\documentclass[journal]{new-aiaa}
\usepackage[utf8]{inputenc}
\usepackage{textcomp}
\usepackage{graphicx}
\usepackage{amsmath}
\usepackage[version=4]{mhchem}
\usepackage{siunitx}
\usepackage{longtable,tabularx}
\usepackage[T1]{fontenc}
\usepackage{caption}
\usepackage{subcaption}
\usepackage{wasysym}
\title{Transfers between moons with escape and capture patterns via Lyapunov exponent maps}

\author{David Canales \footnote[1]{Ph.D. Candidate, Purdue University; Currently, Assistant Professor in the department of Aerospace Engineering in Embry--Riddle Aeronautical University, AIAA Member}$^1$, Kathleen C. Howell\footnote[2]{Hsu Lo Distinguished Professor of Aeronautics and Astronautics, School of Aeronautics and Astronautics, AIAA Fellow}$^2$, Elena Fantino\footnote[3]{Associate Professor, Aerospace Engineering Department, AIAA Member}$^{3}$, Annika J. Gilliam\footnote[4]{Undergraduate Research Assistant, Aerospace Engineering Department}}

\affil[1]{Embry--Riddle Aeronautical University, Daytona Beach, FL, 32117}
\affil[2]{Purdue University, West Lafayette, IN, 47907}
\affil[3]{Khalifa University of Science and Technology, Abu Dhabi, P.O. Box 127788, United Arab Emirates}

\begin{document}

\maketitle

\begin{abstract}
This contribution focuses on the design of low--energy transfers between planetary moons and presents an efficient technique to compute trajectories characterized by desirable behaviors in the vicinities of the departure and destination bodies. The method utilizes finite--time Lyapunov exponent maps in combination with the Moon--to--Moon Analytical Transfer (MMAT) method previously proposed by the authors. The integration of these two components facilitates the design of direct transfers between moons within the context of the circular restricted three--body problem, and allows the inclusion of a variety of trajectory patterns, such as captures, landings, transits and takeoffs, at the two ends of a transfer. The foundations and properties of the technique are illustrated through an application based on impulsive direct transfers between Ganymede and Europa. However, the methodology can be employed to assist in the design of more complex mission scenarios, such as moon tours.
\end{abstract}

\section*{Nomenclature}

{\renewcommand\arraystretch{1.0}
\noindent\begin{longtable*}{@{}l @{\quad=\quad} l@{}}
$\Delta{v}$  & magnitude of velocity variation \\
$t_{tot}$ & total time--of--flight for a transfer \\
$\delta$ & perturbation \\
$\textbf{x}$ & six--dimensional vector accounting for three position and three velocity components\\
$\phi(t_f,t_0)$ & state transition matrix (STM) between $t_f$ (final time) and $t_0$ (initial time)\\
$n$ & integer used to enumerate, e.g., $n$ = 1, 2 \\
$m_p$ & mass of the planet \\
$m_m$ & mass of the moon \\
$\textbf{r}_{p-s/c}$ & position vector of the spacecraft relative to the planet\\
${r}_{p-s/c}$ & distance between the planet and the spacecraft \\
$\textbf{r}_{m-s/c}$ & position vector of the spacecraft relative to the moon \\
${r}_{m-s/c}$ & distance between the moon and the spacecraft \\
$\textbf{r}_{p}$ & position vector of the planet relative to the origin of the reference frame\\
$\textbf{r}_{m}$ & position vector of the moon relative to the origin of the reference frame\\
$\theta$ & true anomaly\\
$t$ & time \\  
CR3BP &  circular restricted three--body problem \\
$\mu$ & mass ratio in the CR3BP  \\
$e$ & orbital eccentricity\\
$a$ & orbital semi--major axis \\
$i$ & orbital inclination \\
$\Omega$ & right ascension of the ascending node of an orbit\\
$\omega$ & argument of periapsis of an orbit\\
${L_a}^*$ & reference length for normalization \\   
$\textbf{r}$ & position of the spacecraft in the planet--moon barycentric rotating reference frame \\
$\dot{\textbf{r}}$ & velocity  of the spacecraft in the planet--moon barycentric rotating reference frame \\
$U^*$ & pseudo--potential function in the CR3BP \\
$J$ & Jacobi constant \\
$L_1, L_2, \dots, L_5$ & five equilibrium points of the CR3BP \\
$x$ & coordinate on the $\hat{x}$--axis of the rotating reference frame \\
$y$ & coordinate on the $\hat{y}$--axis of the rotating reference frame \\
$z$ & coordinate on the $\hat{z}$--axis of the rotating reference frame \\
$\lambda$ & eigenvalue \\
$\Sigma$ & hyperplane associated to a Poincar\'e section \\
$\theta_{0_{m}}$ & true anomaly of the orbit of a moon  measured from the ascending node at the initial epoch\\
$R_{SoI}$ & radius of the sphere of influence of a moon \\
$d_{SoI}$ & ratio  between the gravitational acceleration of the moon and that of the planet \\
$\theta_{Int}$ & true anomaly of the point of intersection between two confocal conic sections\\
$\iota$ & singular values of a matrix \\
$\Upsilon$ & diagonal matrix incorporating all singular values \\
$V$ &  matrix that incorporates the direction of stretching at $t_0$ \\
$U$ &  matrix that incorporates the direction of stretching at $t_f$ \\
$C$ & Cauchy--Green Strain Tensor matrix \\

\end{longtable*}}

\section{Introduction}
\textbf{The} recently released Decadal Strategy for Planetary Science and Astrobiology 2023--2032 \cite{Decadal2022} prioritizes new missions to the gas giants and their moons (e.g., Enceladus multiple flyby and lander, Saturn probe, Titan orbiter, Europa lander, Neptune--Triton probe, Uranus orbiter) and firmly recommends the realization of NASA's Europa Clipper \cite{Clipper2014}. The scientific community believes that the open questions regarding the Sun's planetary system can only be addressed through a systematic exploration of the icy worlds, with particular emphasis on the {\it in situ} observation of planetary moons. In this scenario, the development of efficient tools to design trajectories enabling the execution of transfers between moons and tours of planetary systems is crucial.

Conventional trajectory design methods  based on patched conics and multi--gravity assist have been extensively applied to real mission scenarios (e.g., JUpiter ICy moons Explorer -- JUICE \cite{JUICE2014}, JIMO \cite{JIMO2006}) and have been the focus of numerous studies (see, e.g., \cite{Ross2003,Izzo2013,Colasurdo2014}).
Over the past two decades, investigations in terms of dynamical systems have demonstrated that it is possible to fly a spacecraft on low--energy trajectories departing from and leading to the vicinity of the libration points in the circular restricted three--body problems (CR3BPs) composed of a planet and individual moons. The Petit Grand Tour (PGT) has been the first concept of a low--energy tour of the Jovian system \cite{PGT2001,Gomez2004}. Here,  hyperbolic invariant manifolds of libration point orbits (LPOs) in the CR3BPs associated with Jupiter and distinct moons are propagated in the space between the moons, and their intersections are used to design direct impulsive transfers; a trajectory between LPOs in the vicinity of Ganymede and Europa costs 1.214 km/s and takes 25 days. In the Multi-Moon Orbiter (MMO) concept \cite{Koon2000,Koon2002,Koon2011}, the spacecraft executes several resonant gravity assists with the moons, reducing drastically the propellant consumption to tens of m/s at the expense of increased times of flight (several years). Grover and Ross \cite{Grover2009} employed the Keplerian map and mitigated the long transfer times of the MMO through the introduction of {\it ad hoc} small impulsive maneuvers (summing to $\Delta v$s of 100 m/s). 
The investigation by Campagnola and Russell \cite{Endgame1} on $V_{\infty}$-Leveraging maneuvers (VILMs) led to the identification of Ganymede--to--Europa transfers including endgames (the departure and arrival conditions are low circular orbits) with a total cost of 1.71 km/s (of which 1.41 km/s resulted from achieving escape and capture) and a minimum time of flight of 151 days. The study of moon tours culminated with the blending of resonance hopping transfers and multi-body dynamics with different types of begin and end games, such as circular orbits about the departure and arrival moons ($\Delta v$ = 1.25 km/s and time of flight of 300 days for the Ganymede--to-Europa transfer) \cite{Campagnola2010}, general low-energy initial and final states (59.5 m/s and 158.5 days) \cite{Lantoine2011b}, and halo orbits near the collinear libration points in the vicinity of the two moons (55 m/s  and 205 days) \cite{Lantoine2011a}. 

Following up on the direct transfers developed within the PGT, Fantino and Castelli \cite{Fantino2017} introduced a patched two--body/three-body (2BP-CR3BP) model to facilitate the design of minimum--cost single-impulse moon--to--moon trajectories in the Jovian system using invariant manifolds of planar Lyapunov orbits in two dimensions (2D).  A preliminary extension to trajectories between three-dimensional (3D) halo orbits (Fantino et al. \cite{Fantino2018}) was completed by Canales et al. through the development of an analytical method, termed the Moon--to--Moon Analytical Transfer (MMAT) technique, to construct impulsive transfers between 2D and 3D LPOs of planet--moon CR3BPs \cite{CanalesAAS2020,CanalesCMDA2021,CanalesAA2022a}. 
Invariant manifold trajectories emanating from a departure and a destination LPO are propagated in the respective CR3BPs to the limit of the sphere of influence for the respective moon, where the states of the spacecraft are expressed in a planet--centered inertial frame and used to produce orbital elements of osculating Keplerian orbits. Thus, the problem of connecting trajectories originating from or leading to distinct moons translates into the analytical computation of the intersection between confocal ellipses, the derivation of the conditions under which such intersections exist, and the evaluation of the transfer performance in terms of cost and time of flight. The method incorporates the inclination of the moon orbits, and, in addition to single--impulse transfers, can solve problems with intermediate arcs (two-- and three-impulse scenarios) and plane--change maneuvers in a variety of systems, including trajectories between the Martian moons \cite{CanalesAAS2021,CanalesAA2022b}. 

For single-impulse direct trajectories between planar Lyapunov orbits at Ganymede and Europa, the MMAT approach yields a $\Delta v$  of 0.94 km/s \cite{CanalesCMDA2021}, consistent with the results available in the open literature for these types of trajectories. The predicted time of flight is 9.5 days.
In its original formulation, MMAT deals only with the escape and capture phases of a transfer, and does not resolve the initial and final portions for departing and inserting into the science orbit around each moon, i.e., the so-called begin game/end game problem mentioned previously. 
The objective of the present contribution is to lay the foundations of a strategy to link the inter-moon transfer and the end game design through some desired trajectory patterns. The methodology is applicable within the context of impulsive direct trajectories (MMAT scenarios) as well as  more efficient and practical moon--tour design methods (such as  resonance hoppings  \cite{Campagnola2010,Lantoine2011b,Lantoine2011a}). 

With the aid of chaos indicators, the trajectories that depart or approach the vicinity of a moon can be classified in terms of the motion patterns that they exhibit in close proximity to the target. In particular, it is possible to discriminate among temporary captures (with one or more revolutions around the moon), escapes, takeoffs and impacts. 
The first use of chaos indicators for spacecraft trajectory design in multi-body environments is due to 
Lara et al. \cite{Lara2007} and Villac \cite{Villac2008}, who employed the Fast Lyapunov Indicator, well-known in dynamical astronomy.
In this work, the properties of finite--time Lyapunov exponents (FTLEs) \cite{Haller2011a} and their associated scalar fields that measure the largest stretching direction in the flow associated with the dynamical differential equations are utilized to distinguish phase--space regions corresponding to distinct motion patterns. Low--energy transfers with specific departure and arrival behaviors are designed by coupling FTLE maps and the MMAT method. 
The theoretical foundations, the properties and the advantages of the technique are illustrated through the classical case of direct single-impulse transfers between Ganymede and Europa.

The article is organized as follows. Section~\ref{sec:background} presents the dynamical model and summarizes the background on MMAT and FTLEs. 
Section~\ref{sec:MMATAccessMaps} delves into the basic components of the method, i.e., the moon--to--moon access map, that facilitates the selection of trajectory patterns at the beginning and end of a transfer.  Sections~\ref{sec:access} to \ref{sec:relation} include an analysis of the dependence of the outcome on parameters such as the Jacobi constant and the departure epoch, whereas Section~\ref{sec:inspectionmaps} describes the choice of specific motion patterns on the basis of cost, time--of--flight and departure dates through appropriate inspection maps. Section~\ref{sec:additional} compares inward and outward transfers, and Section~\ref{sec:conclusion} delivers the concluding remarks. 

A preliminary version of the study has been presented by Canales et al. \cite{Canales2021Summer}, whereas the application of the technique to a transfer involving three moons (Io, Europa and Ganymede) has been illustrated in \cite{CanalesASCEND2021}.  

%In this work,  The technique incorporates the true orbital planes of the moons and is applicable to any planetary system. In the vicinity of the moons, the dynamical structures of the CR3BP are leveraged. By establishing a relationship between FTLE maps and the MMAT method, a design technique is developed that computes direct transfers between moons involving captures, collisions or transit orbits (FTLE maps) and characterized by analytical solutions for the $\Delta v$ budget, the \textbf{time--of--flight} and the required relative orbital phases between the moons at departure (MMAT method). This analysis is introduced by Canales, Howell \& Fantino in \cite{Canales2021Summer}, whereas the application of the technique to a transfer involving three moons (Io, Europa and Ganymede) is described in \cite{CanalesASCEND2021}. Here, FTLE maps are introduced into the methodology and blended into the design process. The emphasis is on the theoretical foundations, properties and advantages of the {\bf approach} for designing \textbf{moon--to--moon} transfers with different endgames. Here, FTLE maps are introduced into the methodology and blended into the design process. The emphasis is on the theoretical foundations, properties and advantages of the {\bf approach} for designing \textbf{moon--to--moon} transfers with different endgames.

\section{\label{sec:background} Background and methodology}
After defining the dynamical model, this section provides a summary of the MMAT method and an overview of FTLEs and FTLE maps. The discussion employs direct single-impulse transfers from Ganymede to Europa as applications to illustrate the methodology and its features.

\subsection{Dynamical model} 
\begin{figure}[b!]
\centering
\includegraphics[width=0.6\linewidth]{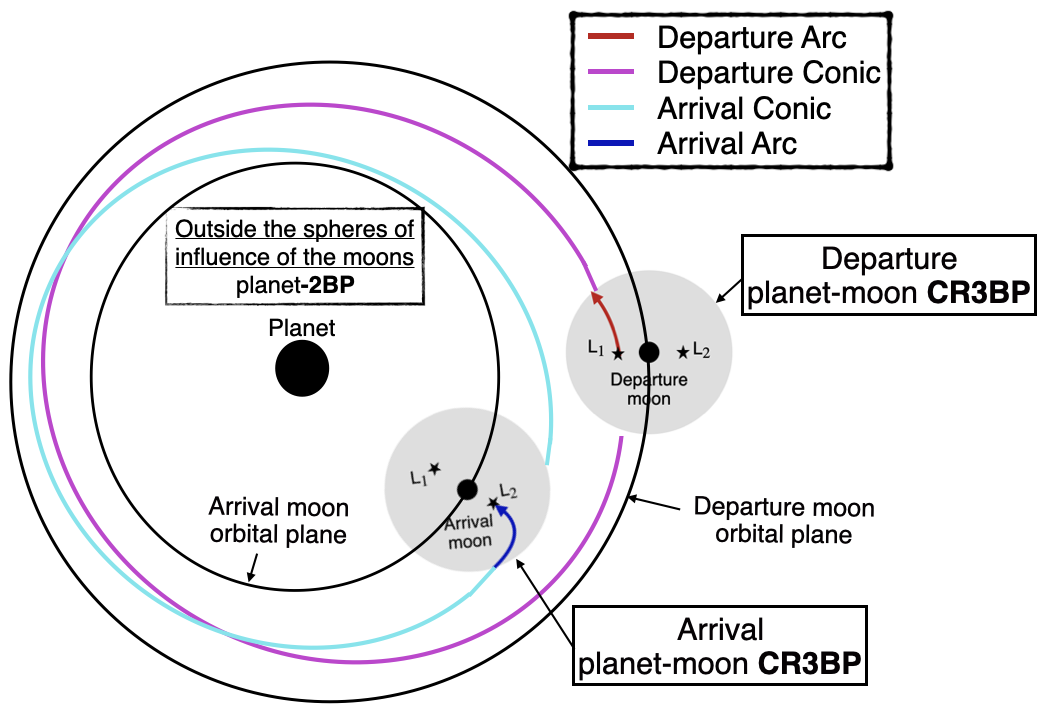}
\caption{Sketch representing the spatial \textbf{2BP--CR3BP} patched model.}
\label{fig:dynamicalmodel}
\end{figure}
In a transfer between moons, the spacecraft  is subject to the gravitational attraction of multiple bodies.
In the CR3BP \cite{Poincare1892}, only two masses (in this case, the planet and one moon) affect the motion of the spacecraft. This approximation is not sufficient to obtain accurate moon--to--moon trajectories, but the results can be refined in a high--fidelity full-ephemeris model.  In this investigation, a spatial 2BP--CR3BP patched model (Fig.~\ref{fig:dynamicalmodel}) is adopted: in the vicinity of each moon, the motion is modeled in the CR3BP of the corresponding planet--moon system; far from the moon, the dynamics of the spacecraft are approximated with the planet--s/c 2BP problem, where the real inclinations of the orbital planes of the moons are incorporated.

The CR3BP provides an appropriate and convenient framework to study the spacecraft motion in the vicinity of a moon. In this model, the planet  (mass $m_p$) and the moon (mass $m_m$) move in circular orbits around the center of mass of the system. Although they exert gravitational attraction upon the spacecraft, the latter does not affect their motion. Additionally, the primaries are spherical and their mass distribution is homogeneous, i.e., gravity field irregularities are disregarded. A suitable normalization of distances, masses and angular velocities and the adoption of an appropriate barycentric rotating reference frame leads to
the following set of dimensionless differential equations for the motion of the spacecraft \cite{Szebehely1974}:
\begin{align}
\label{eq:EOM}
\ddot{x} - 2\dot{y} = \frac{\delta{U^*}}{\delta{x}};\;\;\;  \ddot{y} + 2\dot{x} = \frac{\delta{U^*}}{\delta{y}};\;\;\; \ddot{z} = \frac{\delta{U^*}}{\delta{z}}.
\end{align}
Here, $\mu = \frac{m_m}{(m_m+m_p)}$ is the mass ratio of the system, whereas the term $U^* = \frac{1-\mu}{r_{p-s/c}} + \frac{\mu}{r_{m-s/c}}+\frac{1}{2}(x^2+y^2)$ represents the pseudo--potential function, $r_{p-s/c}$ and $r_{m-s/c}$ being the distances of the spacecraft to the planet and the moon, respectively.
The $\hat{x}$--axis of this rotating reference frame contains both primaries (the planet at $\textbf{r}_p = [-\mu,0,0]^T$ and the moon at $\textbf{r}_m = [1-\mu,0,0]^T$), whereas the $\hat{z}$--axis is aligned with their orbital angular momentum. 
In Eq.~\eqref{eq:EOM}, $\textbf{r}_{rot} = [x,y,z]^T$ and $\dot{\textbf{r}}_{rot} = [\dot{x},\dot{y},\dot{z}]^T$ are the position and the velocity of the spacecraft, respectively. The CR3BP admits five equilibrium positions (denoted as libration points and labelled $L_1$, $L_2$, ..., $L_5$) i.e., points where the acceleration is zero if the third body is at rest. As the libration points exhibit linear stability properties, orbits around them are categorized by families of periodic and quasi--periodic orbits \cite{brouckePOs1968,Campbell1999}. The planar Lyapunov orbit family is of interest in this work. In the vicinity of a moon, hyperbolic invariant manifolds extending from periodic orbits operate as pathways, including connections to other periodic orbits within the same system \cite{Haapala2015}. Additionally, stable manifolds arrive near a periodic orbit while unstable manifolds depart from its vicinity. Transit orbits, as defined for this investigation, reach the moon vicinity through the $L_1$ and $L_2$ gateways defined by periodic orbits, revolve around the moon and either depart again or collide with it.  The Jacobi constant $J$, that represents the energy of the system, is a fundamental quantity in the CR3BP. It is defined as
\begin{align}
\label{eq:JC}
J = 2U^* - (\dot{x}^2+\dot{y}^2+\dot{z}^2).
\end{align}
The model in the CR3BP is well-known and serves as the foundational framework in multi-body regimes.

\subsection{The MMAT method} 
\begin{figure}[b!]
\centering
\includegraphics[width=0.75\linewidth]{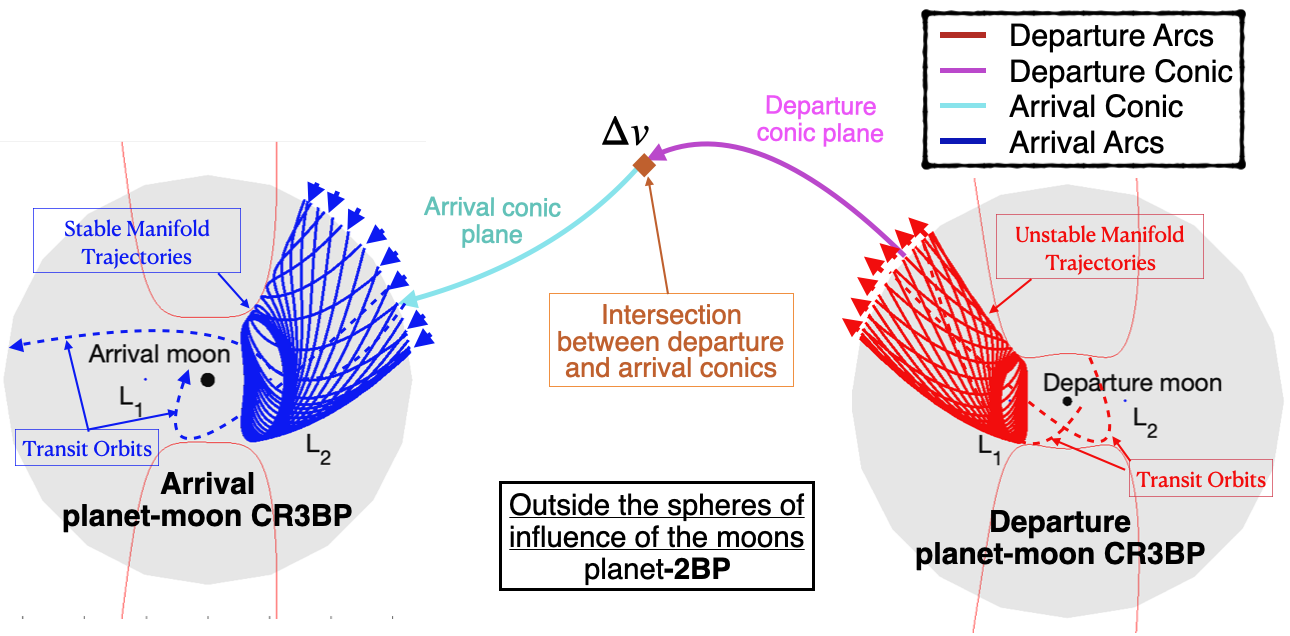}
\caption{Scheme illustrating the MMAT method for constructing direct transfers between moons.}
\label{fig:directtransfer}
\end{figure}
In the patched 2BP-CR3BP, within the Sphere of Influence (SoI) of the moon the motion is modeled in the planet-moon CR3BP. When the trajectories cross the surface of the SoI, they are approximated in the planet--s/c 2BP as conic sections with one focus at the planet. Therefore, the motion outside the SoI is considered Keplerian and completely determined by six osculating orbital elements computed at the surface of the SoI. In this investigation, the semi--major axis ($a$), the eccentricity ($e$) and the true anomaly ($\theta$) are key parameters for moon--to--moon transfer design. In this patched 2BP--CR3BP model, connections between conics that depart from and arrive at distinct moons are explored through analytical methods. The SoI of a moon is defined as a spherical region centered at the moon with radius $R_{SoI}$ equal to the distance from the moon along the $x$--axis at which the ratio ($d_{SoI}$) between the gravitational acceleration due to the moon and that caused by the planet equals a certain small quantity, which is a free parameter. The value adopted for this investigation for the SoI of both Ganymede and Europa is $d_{SoI}=5\times10^{-4}$. The selection of this value affects the design of the moon--to--moon transfer and is examined for this particular application by Canales et al. \cite{CanalesCMDA2021}.

%The objective of this investigation is to design \textbf{moon--to--moon} transfers in which the departure and arrival trajectories are solutions in the respective \textbf{planet--moon} CR3BPs. 
This study incorporates the inclinations of the orbital planes of the moons and, therefore, the analysis is developed in three-dimensions (3D). Although there exist many possible departure and arrival paths, this 3D nature makes the problem dependent on the relative orbital phase between the moons at the departure epoch. The MMAT method (see Fig.~\ref{fig:directtransfer}) previously presented by the authors \cite{CanalesCMDA2021} utilizes the patched 2BP--CR3BP: the trajectories arriving or departing a moon vicinity are represented using conic arcs outside the moons' SoI. In a transfer from Ganymede to Europa, departure conics approximate the departure Jupiter--Ganymede (J--G) CR3BP orbits, while arrival conics approximate the arrival Jupiter--Europa (J--E) CR3BP orbits. The approximation through planet--centered Keplerian orbits neglects the gravitational attraction of the moons, but the associated error is small (see  \cite{Fantino2017}) and the solutions can be efficiently transitioned to a high--fidelity ephemeris model.

The necessary condition for an arrival conic to intersect spatially with a departure conic can be expressed analytically as:
\begin{align}
\label{eq:4.15fromdissertation}
a_a(1-e_a)\le\frac{a_d(1-{e_d}^2)}{1+{e_d}\cos(\theta_{d_{Int}}+n\pi)}\le{a_a(1+e_a)}, \;\; \text{with } n = 0,1.
\end{align}
Here, $a_a$ and $a_d$ are the semi--major axes of the arrival and departure conics, respectively. Similarly, $e_a$ and $e_d$ are the respective eccentricities. The true anomalies $\theta_{d_{Int}}$ and $\theta_{d_{Int}} + \pi$ indicate the two geometrical configurations for which intersections between the planes of the departure and arrival conics exist. For a given departure epoch, if the condition in Eq.~\eqref{eq:4.15fromdissertation} is satisfied, a suitable orbital phase for the target arrival moon can be determined. The correct phasing for the two moons yields a transfer between the conic arcs through a single impulsive $\Delta{v}$. Each possible transfer is characterized by a total time--of--flight $t_{tot}$ and requires an impulsive maneuver of magnitude $\Delta{v}$. In summary, given the departure epoch from one moon, the MMAT technique identifies possible arrival conditions and moon--to--moon transfers with one impulse and different performance characteristics.

\subsection{Finite--time Lyapunov exponents} 
\begin{figure}[b!]
\centering
\includegraphics[width=0.5\linewidth]{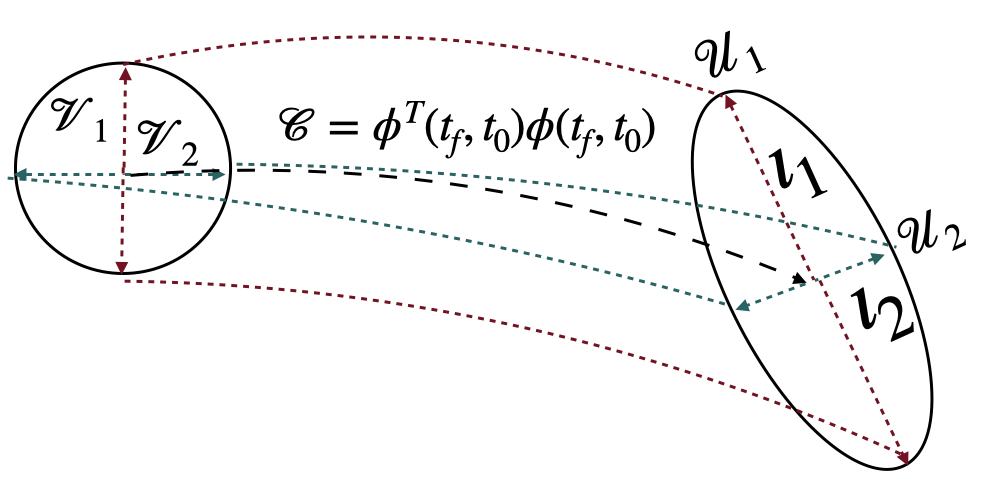}
\caption{\label{fig: stretching} Cauchy--Green Strain Tensor associated eigenvector stretching.}
\end{figure}

The concept of FTLEs is related to that of Lagrangian Coherent Structures (LCS) \cite{Wiggins1992,Gawlik2009,Short2014,PerezPalau2012,Haller2011b}, i.e., regions bounding different behaviors in a dynamical flow.
Dynamical systems theory leverages different techniques with two--dimensional Poincar\'e maps to investigate the long--term dynamics in the CR3BP. The Cauchy--Green Strain Tensor (CGST) \cite{CGST19971} effectively describes the time evolution of the flow resulting from a perturbation. To introduce the CGST, the state transition matrix (STM) is required. The STM provides a relationship between a variation of the initial state $\delta{\textbf{x}_0}$ and the resulting deviation of the final state $\delta{\textbf{x}_f}$ via the linear mapping:
\begin{align}
\label{eq:STM}
\delta\textbf{x}_f = \phi(t_f,t_0)\delta\textbf{x}_0.
\end{align}
The variables $t_0$ and $t_f$ represent the initial and final times along the trajectory, respectively, whereas  $\phi(t_f,t_0)$ is the STM. Through a singular--value decomposition (SVD) \cite{doi:10.1137/0702016} of the STM, it is possible to characterize a perturbation behavior as divergence or convergence. Such characteristics of the local phase space are encoded in singular values ($\iota_i$) defined by the direction of stretching or contraction $\textbf{V}_i$ at $t_0$. The SVD is expressed as:
\begin{align}
\label{eq:singularvaluedecomp}
\phi(t_f,t_0) = \boldsymbol{U}\Upsilon{\boldsymbol{V}^T}.
\end{align}
Here boldface letters represent matrices. Then, $\boldsymbol{U}$ and $\boldsymbol{V}$ are mutually orthogonal: $\boldsymbol{V}$ identifies the directions of stretching at $t_0$ for every singular value, whereas the elements of the columns of $\boldsymbol{U}$ define the directions of stretching (or contraction) at $t_f$. Additionally, $\Upsilon$ is a diagonal matrix representing the stretching magnitude, written with the singular values in descending order ($\iota_1>\iota_2>...>\iota_n$):
\begin{align}
\label{eq:Upsilon}
\Upsilon = 
{\begin{bmatrix}
\iota_1 & 0 & \dots & 0\\
0 & \iota_2 & \dots & 0\\
\vdots & \vdots & \ddots & \vdots\\
0 & 0 & \dots & \iota_n\\
\end{bmatrix}}.
\end{align}
Figure~\ref{fig: stretching} sketches the SVD in a simplified two--dimensional space: in this scheme, $\iota_1$ and $\iota_2$ indicate the smallest and largest stretching, respectively, revealing which directions are more or less sensitive to perturbations, with larger magnitudes indicating higher sensitivity. Once the STM is defined, the CGST describes the deformation of the flow as the product of the STM and its transpose \cite{Smith1993}:
\begin{align}
\label{eq:CGST}
\boldsymbol{C}(t_f,t_0) = \phi^T(t_f,t_0)\phi(t_f,t_0).
\end{align}
The eigenvalues ($\lambda$) of $\boldsymbol{C}$ are obtained from the eigen--decomposition of the CGST, and are related to $\Upsilon$ through $\lambda_i = {\iota_i}^2$. Additionally, $\boldsymbol{C}$ and $\boldsymbol{U}$ possess the same eigenvectors. Given that the largest singular value corresponds to the largest perturbation growth, the FTLEs are defined based on $\iota_1$ and the propagation time along the trajectory: 
\begin{align}
\label{eq:FTLE}
\text{FTLE} = \frac{\iota_1}{|\Delta{t}|}.
\end{align}
Here $\Delta{t} = t_f-t_0$. Hence, all stretching directions produce an FTLE, but only the direction of maximum stretching is employed in Eq.~\eqref{eq:FTLE}. This value is the one of interest for the definition of LCS. In summary, FTLEs measure the relative phase--space element growth and contraction over a time interval related to the system flow. 

\subsection{FTLE maps}
The benefits of FTLEs emerge when paired with Poincar\'e section representations. Finite--time Lyapunov exponent maps have been proven useful by various authors (see, e.g., \cite{Short2014}) to provide quantitative information about the propagation of a trajectory and the type of motion that occurs at a given energy level after departure or before arrival at a moon vicinity. Possible behaviors that are discerned include capture orbits, collision paths (or landings),  departure trajectories (or takeoffs)  and  close passages by the moons (or transits). Additionally, FTLE maps illustrate separated flow patterns entering the region of the moon through the corresponding gateway (either $L_1$ or $L_2$). They have been recently employed to design transfers that approach Oberon and Titania in the Uranus system and are characterized by a desired behavior \cite{Short2015}. These low--energy trajectories are produced by coupling Uranus--Oberon and Uranus--Titania CR3BPs and assuming that the moons revolve in coplanar orbits. 

\begin{figure}[b!]
\centering
\begin{subfigure}{.45\textwidth}
    \includegraphics[width=1\linewidth]{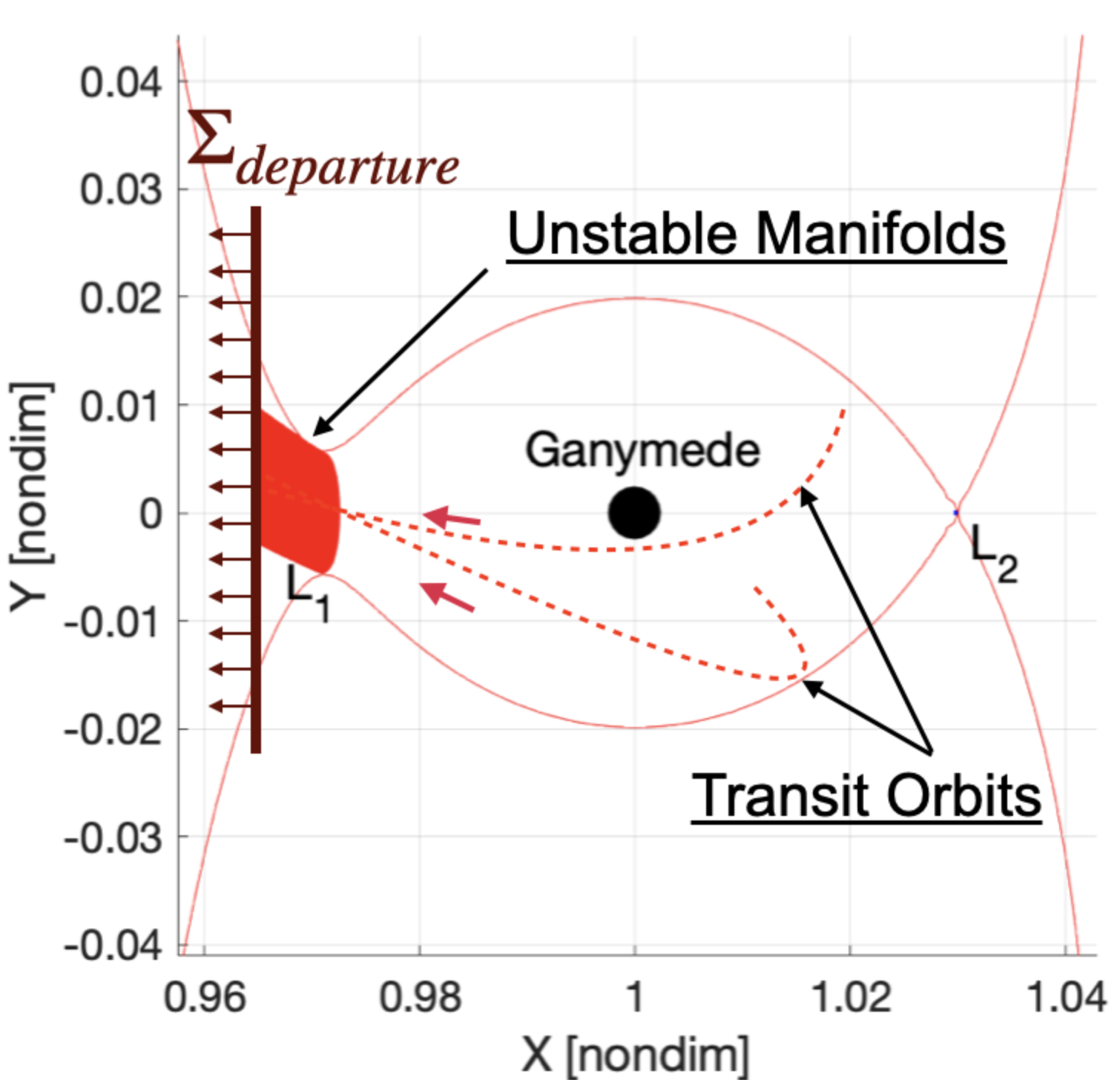}
    \caption{\label{fig:departureScheme} Departure trajectories from Ganymede vicinity through $\boldsymbol{L_1}$ in the J--G rotating frame.}
\end{subfigure} \hspace{2.2cm}
\begin{subfigure}{.4\textwidth}
    \includegraphics[width=1\linewidth]{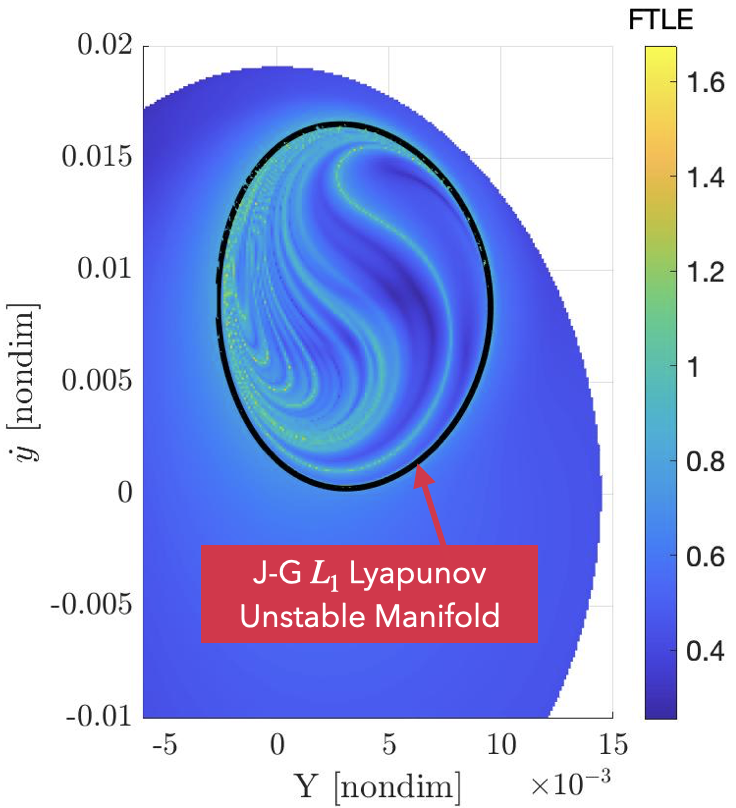}
    \caption{\label{fig:depFTLE} Departure FTLE map at a $\boldsymbol{J_d = 3.007538}$ with a propagation time $\boldsymbol{t_d \approx -11.4}$ days.}
\end{subfigure}

\begin{subfigure}{.44\textwidth}
    \includegraphics[width=1\linewidth]{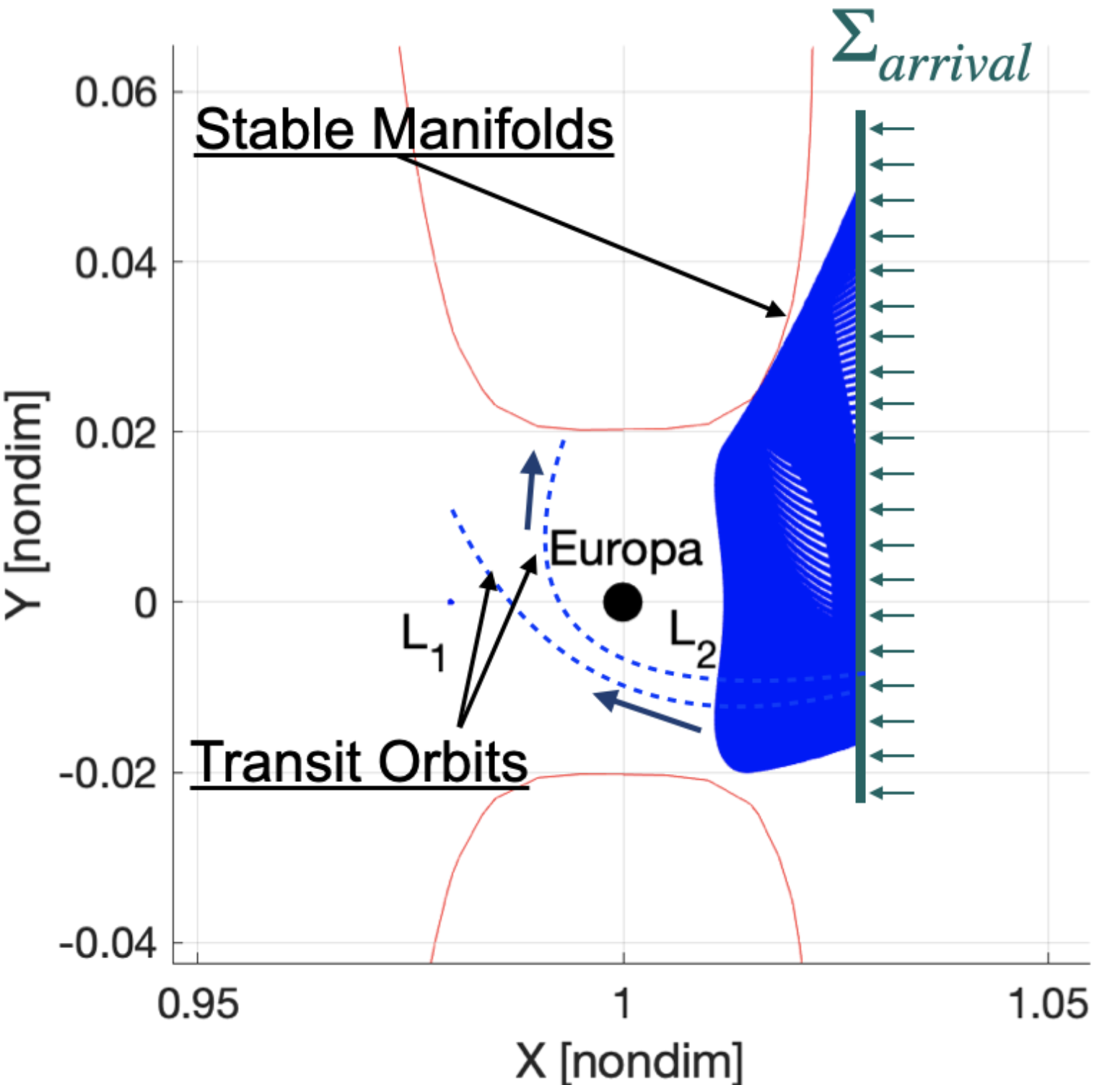}
    \caption{\label{fig:arrivalScheme} Europa arrival trajectories through $\boldsymbol{L_2}$ in the \textbf{J--E} rotating frame.}
\end{subfigure} \hspace{0.5cm}
\begin{subfigure}{.5\textwidth}
    \includegraphics[width=1\linewidth]{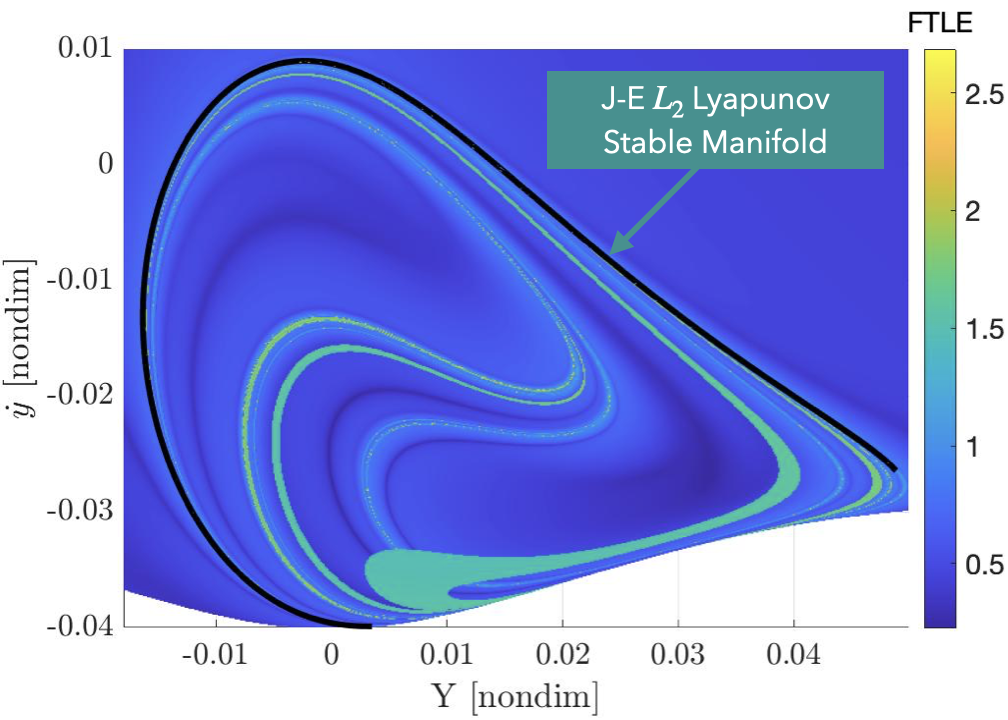}
    \caption{\label{fig:arrFTLE} Arrival FTLE map at a $\boldsymbol{J_a = 3.00240}$ and with propagation time $\boldsymbol{t_a \approx 5.65}$ days.}
\end{subfigure}
\caption{\label{fig:FTLEMapsDep&Arr} (a) Departure trajectories from Ganymede and (b) departure FTLE map; (c) Trajectories arriving at Europa and (d) arrival FTLE map.}
\end{figure}

The computation of  FTLE maps involves a number of parameters, e.g., the location of the Poincar\'e section, the propagation time and the value of the Jacobi constant \cite{Short2014}. To illustrate this methodology, trajectories departing  Ganymede and arriving at Europa are employed (see Table \ref{Table:OrbitalData} for relevant system data).
\begin{table}[t]
\caption{Orbital data for Ganymede and Europa relative to the Ecliptic and Equinox of J2000.0 Jupiter--centered frame \cite{SPICE}.}
\label{Table:OrbitalData}
\centering 
{\footnotesize
\begin{tabular}{lclclclclclclc}
\hline\hline
  Moon    &		\textbf{Semi--major} 	     &	Orbital	 &CR3BP   & Eccentricity & Inclination & Longitude \\
&	axis     & period & mass ratio &   &  & asc. node\\
&	 {[}$10^{5}$ km{]}    &  {[}day{]}& {[}$10^{-5}${]}   &{[}$10^{-3}${]}  & {[}degree{]} &{[}degree{]}\\
\hline
	Europa & $6.713$& 3.554 & 2.528 &  9.170 & 2.1 & 331.4\\
	Ganymede & $10.706$ & 7.158 & 7.804 &  2.542 & 2.2 & 340.3\\
\hline
\end{tabular}
}
\end{table}
The Jacobi constant values are ${J}_d = 3.00754$ and ${J}_a = 3.00240$, respectively in the Ganymede departure and Europa arrival CR3BPs. Separate FTLE maps are created for departure and arrival trajectories (Fig.~\ref{fig:FTLEMapsDep&Arr}). The departure map is based on a departure from Ganymede via a Poincar\'e section near the Jupiter--Ganymede $L_1$ gateway. In Fig.~\ref{fig:FTLEMapsDep&Arr}(a), the section is denoted $\Sigma_{departure}$ and is mapped over an interval of normalized time through the propagation of all the states onto the section. The departure section is created at $x=0.965$ in the J--G rotating frame, where all departure states are generated. The $y$ coordinate varies from $-0.006$ to 0.015, while the range for $\dot{y}$ extends from -0.01  to 0.02. Finally, $\dot{x}$ is computed for each state by means of Eq. \eqref{eq:JC}. Additionally, in this case, the interval $t_d$ equals $-11.4$ days, corresponding to $-10$ normalized J--G CR3BP time units. Note that $t_{tot}$ is a crucial parameter for computing the FTLEs. A numerical analysis has been accomplished to select a value of $t_{tot}$ that provides different regions of interest within the FTLE maps and, thus, different trajectory behaviors over the given time interval. Recall that $t_{tot}$ and the $x$-coordinate of the section are parameters to be selected when computing FTLE maps.

Similarly, the arrival map (Fig.~\ref{fig:FTLEMapsDep&Arr}(d)) is constructed on a Poincar\'e section ($\Sigma_{arrival}$) near the J--E $L_2$ gateway. In Fig.~\ref{fig:FTLEMapsDep&Arr}(c), $\Sigma_{arrival}$ is the plane $x=1.028$ in the J--E rotating frame. A grid is generated by varying the $y$ coordinate from $-0.018$ to 0.05 and $\dot{y}$ from $-0.04$ to 0.01. In this study, the step size adopted to produce the departure and arrival maps is 0.0001 for both coordinates. Using Eq. \eqref{eq:JC}, $\dot{x}$ is  computed for each state. The propagation time interval for arrival is $t_a = 10$ normalized Jupiter--Europa CR3BP time units, equivalent to $5.65$ days.

The selected propagation times for departure and arrival ensure a sufficient emergence of LCS that include different types of behaviors over a considerable time. The FTLE maps depend on the $x$ coordinate of the sections and on the propagation time \cite{phdthesisGarcia}. Note that, in this work, the normalized units in the FTLE maps are those of the planet--moon CR3BP for which the map is being constructed. The trajectories that transit throughout the vicinity of the moons are contained in the hyperbolic invariant manifolds associated with the planar Lyapunov orbits at the given Jacobi constant levels near $L_1$ and $L_2$, respectively (i.e., unstable for J--G CR3BP, stable for the J--E CR3BP). 

\begin{figure}
\hfill{}\centering\includegraphics[width=16cm]{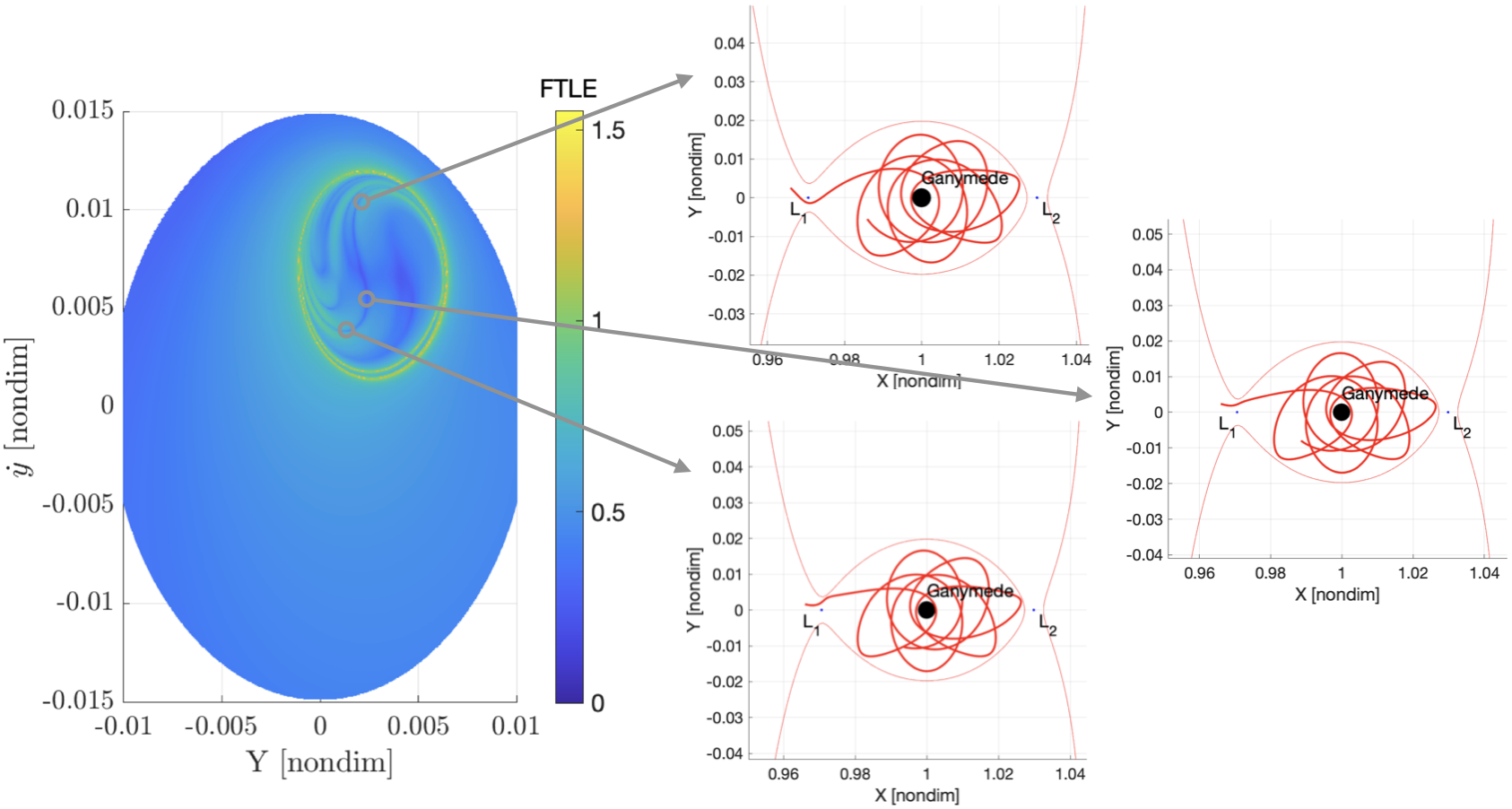}\hfill{}
\caption{\label{fig:streamsWithSameBehavior}Three departure trajectories with the same departure pattern and the same associated isoline upon the FTLE map.}
\end{figure}

\begin{figure}
        \hfill{}\centering\includegraphics[width=13cm]{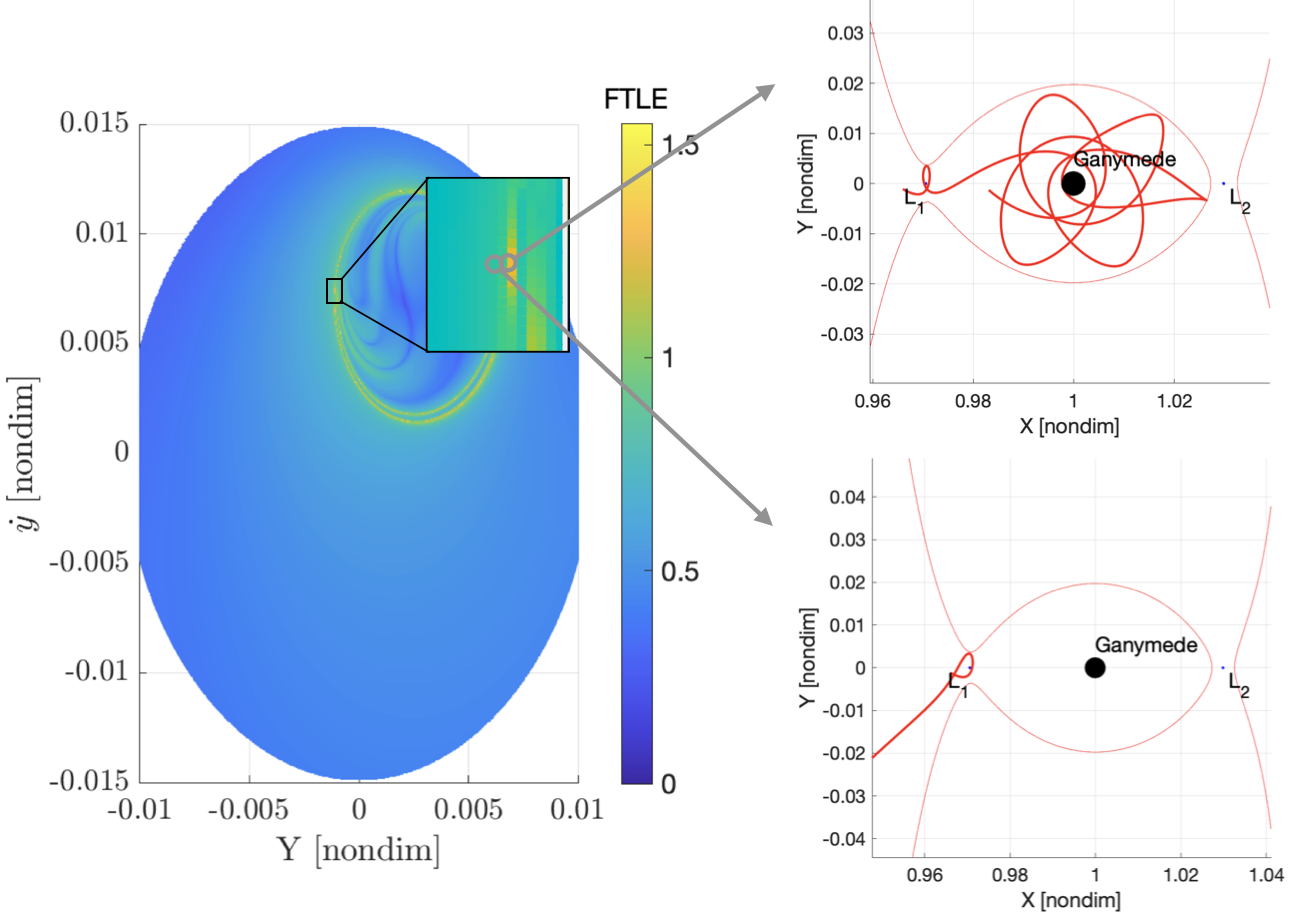}\hfill{}
\caption{\label{fig:streamsLimittingCaptureandnot}Different isolines in the departure FTLE map corresponding  to transit and non--transit trajectories.}
\end{figure}

\begin{figure}
\hfill{}\centering\includegraphics[width=16cm]{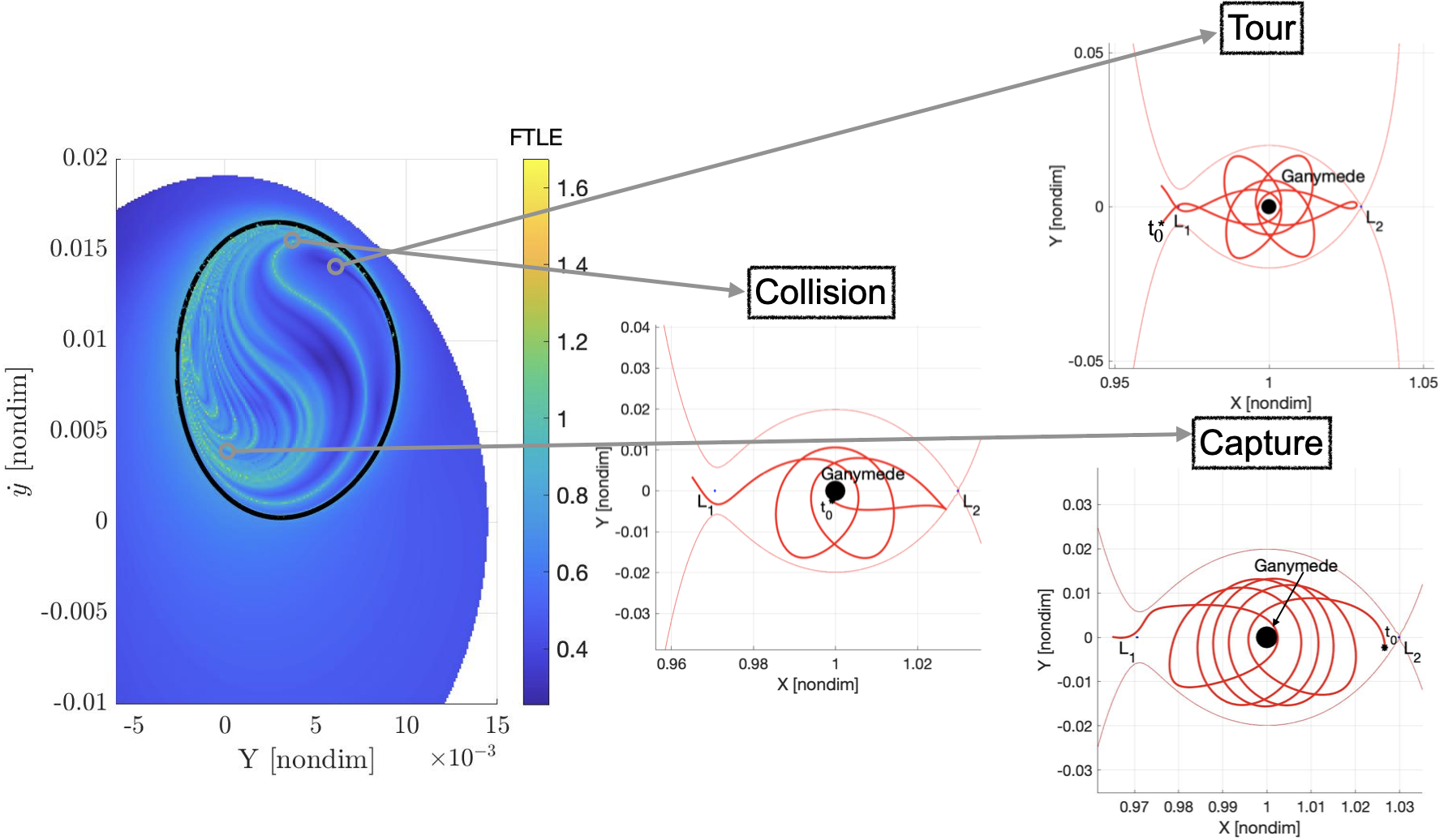}\hfill{}
\caption{\label{fig:diferentCases}Three departure trajectories with different patterns and their distinct isolines on the departure FTLE map ($J_d=3.00754$).}
\end{figure}

As shown in Fig.~\ref{fig:streamsWithSameBehavior}, strainlines bound regions (or ``lobes'') of qualitatively similar motion. This property allows the designer to select initial conditions that follow a desired behavior since all the initial conditions in any such ``lobe'' lead to similar trajectory patterns. As observed in Fig.~\ref{fig:streamsLimittingCaptureandnot}, FTLE maps separate initial conditions corresponding to trajectories that enter the moon's vicinity from those that do not approach the moon. Finally, within the maps themselves, strainlines represent various behaviors and outcomes (see Fig.~\ref{fig:diferentCases}). Hence, they inform in the selection of initial conditions for specified departure and arrival characteristics. Every ``lobe'' in Fig.~\ref{fig:diferentCases} yields a specific motion pattern, such as captures, tours and collisions. Moreover, the separatrices between ``lobes'' identify collision trajectories because when an impact occurs, the  propagation time of the trajectory is relatively short, and this produces large values of the FTLEs (e.g., 1.352 for the collision trajectory in Fig. ~\ref{fig:diferentCases}), resulting in distinctive patterns. Finally, low FTLE values are associated with the centers of the ``lobes''.  This fact is justified since smaller FTLEs (e.g., 0.315 for the selected tour trajectory in Fig. ~\ref{fig:diferentCases}) involve a lower sensitivity to perturbations. It is important to emphasize that the only purpose of the raw FTLE values is to identify distinct ``lobes'' that generate different trajectory behaviors.

\section{\label{sec:MMATAccessMaps}Designing \textbf{moon--to--moon} transfers with different trajectory patterns: the moon--to--moon access maps methodology}
\begin{figure}[b!]
\centering
\includegraphics[width=0.8\linewidth]{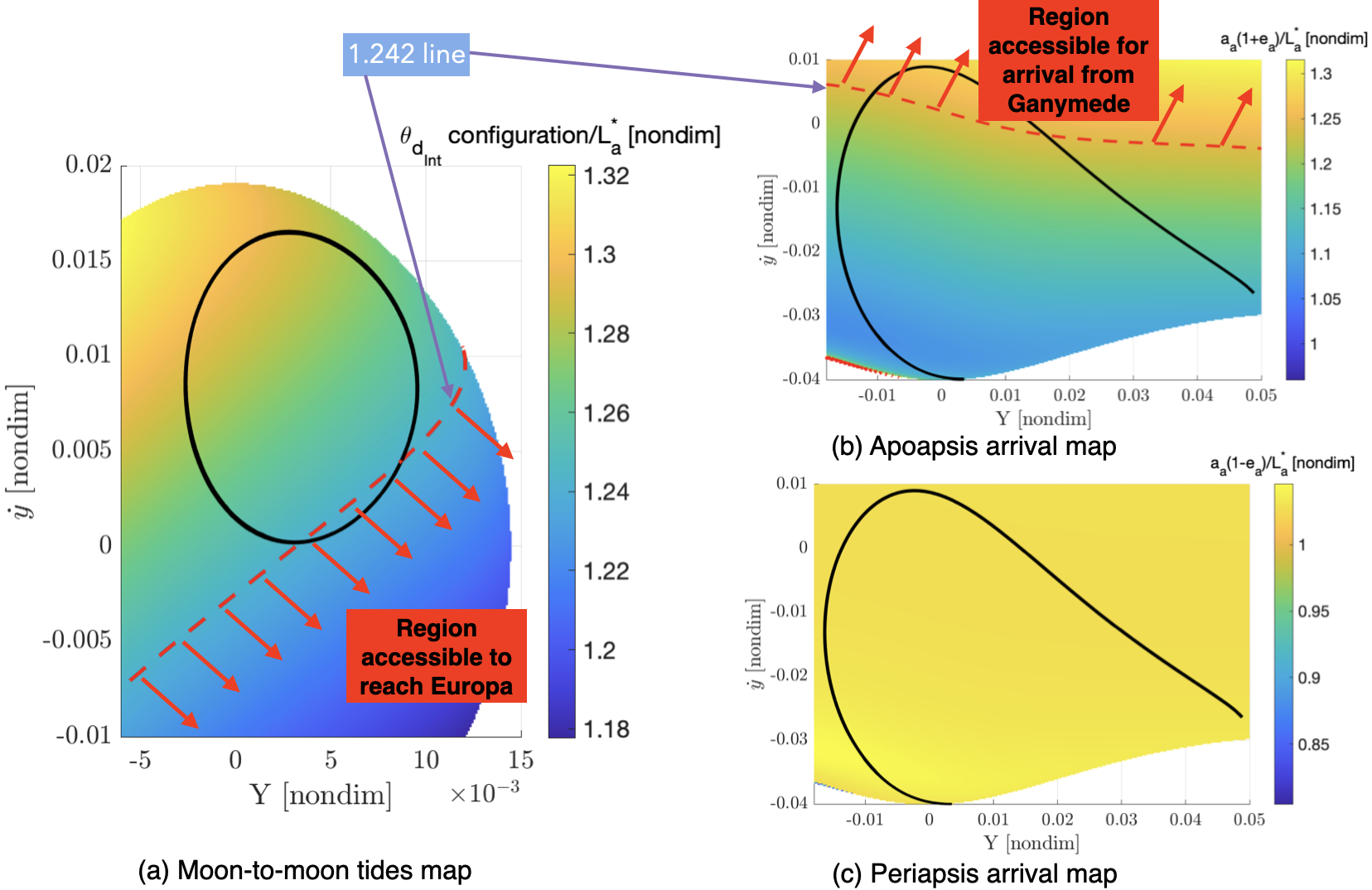}
\caption{Moon--to--moon tides map (a) and apoapsis (b) and periapsis (c) arrival maps matched with 1.242 isoline for transfers from Ganymede towards Europa.}
\label{fig:MMATMAPS}
\end{figure}

\begin{figure}
\centering
\includegraphics[width=0.8\linewidth]{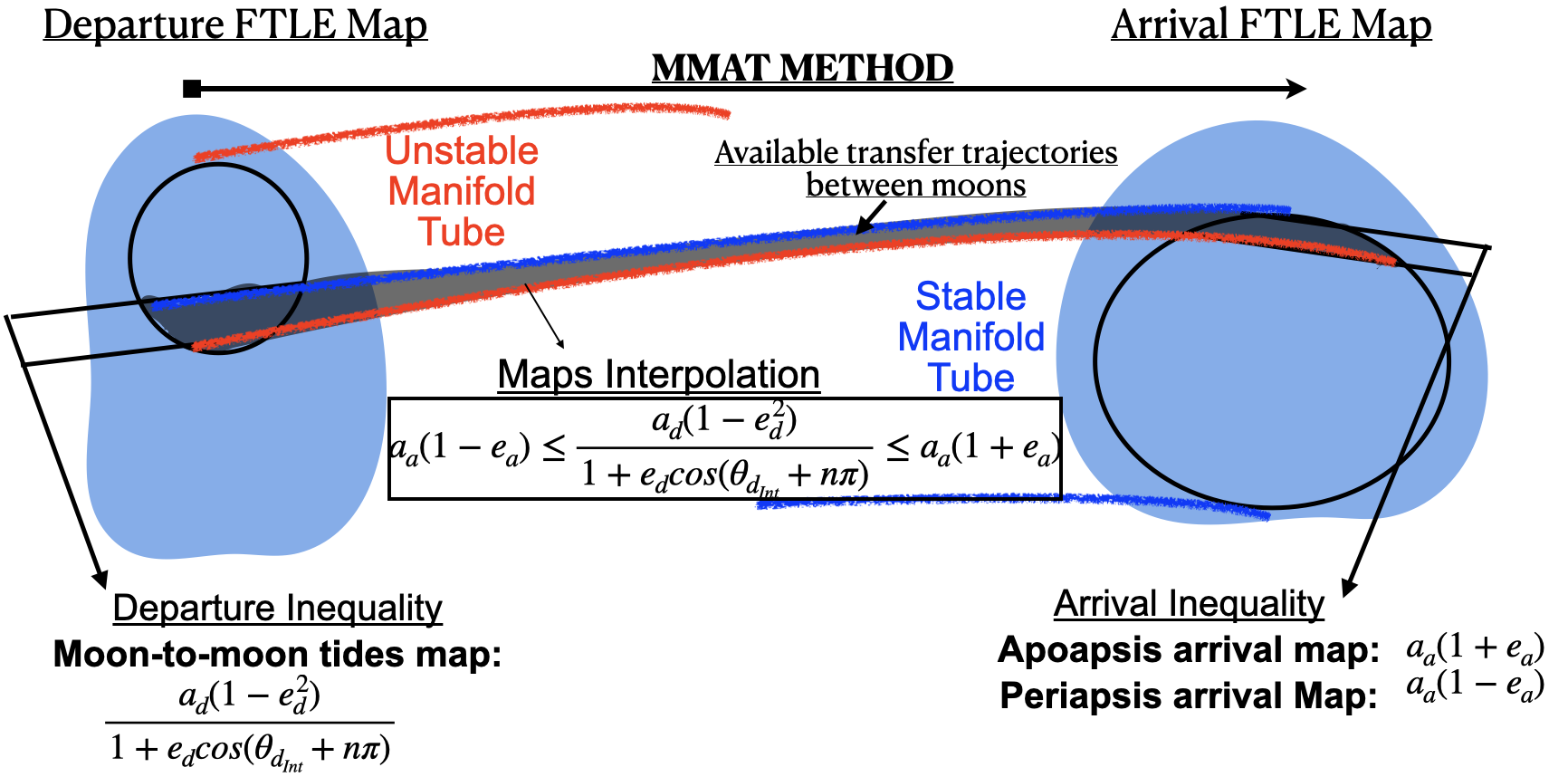}
\caption{Schematic of the moon-to-moon access maps method.}
\label{fig:MMATMAPS2}
\end{figure}

\begin{figure}[]
\hfill{}\centering\includegraphics[width=14cm]{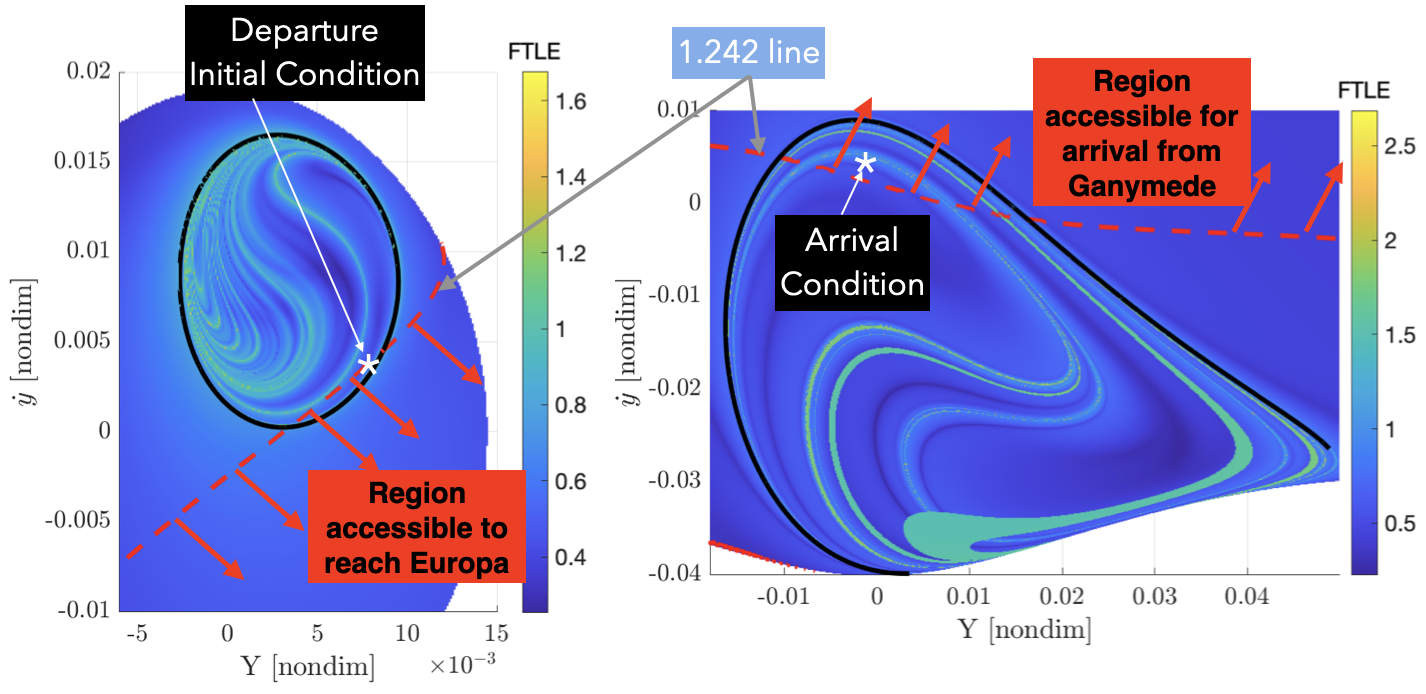}\hfill{}
\caption{\label{fig:availableTransfersFTLEMapsOld} Available trajectories from Ganymede to Europa corresponding to the 1.242 isoline shown through the departure and arrival FTLE maps.}
\end{figure}

\begin{figure}[]
\hfill{}\centering\includegraphics[width=16cm]{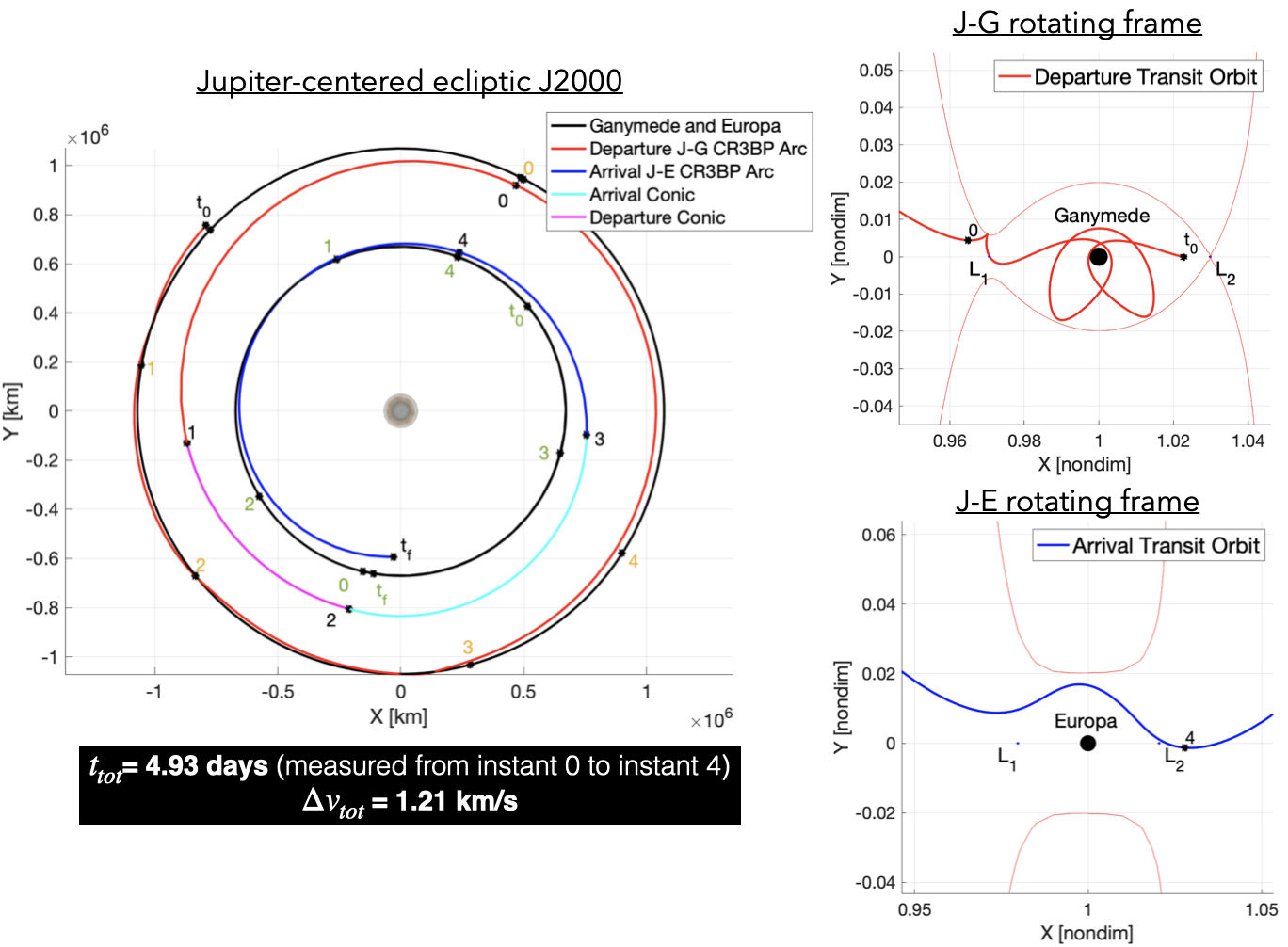}\hfill{}
\caption{\label{fig:exampleTransfer} Transfer from Ganymede to Europa with a desired behavior, as selected in the FTLE maps.}%: \textbf{Jupiter--centered} Ecliptic J2000 frame (left), \textbf{Jupiter--Ganymede} rotating frame (top right) and \textbf{Jupiter--Europa} rotating frame (bottom right).}
\end{figure}

Merging the MMAT method with FTLE maps facilitates the design of moon--to--moon transfers with selected behaviors in the vicinity of the moons. Consider the framework for transfers from Ganymede to Europa in Fig.~\ref{fig:FTLEMapsDep&Arr}. Recall that moon--to--moon transfers are dependent on the departure epoch because the moons revolve in their true orbital planes \cite{CanalesCMDA2021}. 
Equation~\eqref{eq:4.15fromdissertation} must be satisfied for the Ganymede and Europa vicinities to be connected with a single impulsive maneuver. Therefore, the selection of a suitable departure epoch is critical for the identification of effective transfers. 
Assume that all initial conditions depart Ganymede at epoch $\theta_{0_{Gan}} = 82.506^{\circ}$\footnote{Here time
is expressed as the true anomaly of the moon, measured from the ascending node of its orbit.}. Experiments demonstrate that this value 
leads to the largest number of connections between Ganymede and Europa for the specified departure and arrival energy levels (see below). 
Then, since outside the SoIs the trajectories are approximated by conics in the Jupiter--centered 2BP, the central expression in Eq.~\eqref{eq:4.15fromdissertation} is evaluated for every departure conic generated by each initial condition on the Poincar\'e section in the vicinity of the $L_1$ gateway. Note that $\Sigma_{departure}$ lies within the SoI. The map depicted in Fig.~\ref{fig:MMATMAPS}(a) is the moon--to--moon
tides map. Its color gradients reflect the value of the expression $\frac{1}{{L_a}^*}\frac{a_d(1-{e_d}^2)}{1+{e_d}\cos(\theta_{d_{Int}})}$, i.e., the central term of Eq.~\eqref{eq:4.15fromdissertation}, as each initial condition is propagated from the Poincar\'e section to the SoI of Ganymede. This expression is labelled the ``$\theta_{d_{Int}}$ configuration'' and includes the reference length ${L_a}^*$ for normalization in the J--E CR3BP. Observe that the moon--to--moon tides map and the departure FTLE map have the same shape; however, the former uses the propagation of initial conditions towards the SoI and represents the ``$\theta_{d_{Int}}$ configuration'', whereas the latter employs the propagation from the departure moon and illustrates the FTLE. The Europa arrival states are propagated backwards in time from $\Sigma_{arrival}$ towards the SoI of Europa. Then, apoapsis and periapsis arrival maps (Figs.~\ref{fig:MMATMAPS}(b) and (c), respectively) are produced by evaluating the expressions on the left and right sides of Eq.~\eqref{eq:4.15fromdissertation} ($a_{a}(1+e_{a})/L_a^*$ and $a_{a}(1-e_{a})/L_a^*$, respectively) for each arrival condition at the SoI. Note that the reference length for the J--E CR3BP is used to normalize all three maps. The moon--to--moon tides map and the apoapsis and periapsis arrival maps are then matched to identify initial conditions that satisfy Eq.~\eqref{eq:4.15fromdissertation}. For one--impulse transfers to be feasible, the middle term in this equation must be simultaneously $\le a_a(1+e_a)$ (upper constraint) and $\ge a_a(1-e_a)$ (lower constraint). This methodology is denoted the moon--to--moon access maps approach and is sketched in Fig.~\ref{fig:MMATMAPS2}: the gray region corresponds to the Ganymede--to--Europa transfers that satisfy Eq.~\eqref{eq:4.15fromdissertation}.
A specific isoline is selected to accomplish the matching in Fig.~\ref{fig:MMATMAPS}: 
\begin{align}
\label{eq:isoline}
\frac{{a}_d(1-{{e}_d^2)}}{1+{e}_d{\cos({\theta}_{{d}_{Int}}})}={a}_a(1+{e}_a)=1.242 \; {L}_a^*.
\end{align}
For transfers from an outer to an inner moon, the moon--to--moon tides map and the apoapsis arrival map are matched with respect to the requirement $\frac{a_d(1-{e_d}^2)}{1+{e_d}\cos(\theta_{d_{Int}}+n\pi)} \le a_a(1+e_a)$. The uniform color of the periapsis arrival map in Fig.~\ref{fig:MMATMAPS} means that the periapsis radii of all arrival trajectories are similar. Note that since these are smaller than the middle term in Eq.~\eqref{eq:4.15fromdissertation}, the lower constraint is always satisfied for this particular application. Similarly, the upper constraint is always satisfied for a transfer from an inner to an outer moon, and the middle term  of Eq.~\eqref{eq:4.15fromdissertation} must be larger than the arrival conic periapsis radius.
Within the MMAT access maps, red arrows indicate accessible regions, i.e., sets of initial conditions for which Eq.~\eqref{eq:4.15fromdissertation} is fulfilled and the construction of one--impulse trajectories from Ganymede to Europa with different trajectory patterns is possible; in other words, an arrival epoch exists and the transfer is feasible. Projecting the isolines onto the departure and arrival FTLE maps (Fig.~\ref{fig:availableTransfersFTLEMapsOld}) allows inspection for distinct transfers between Ganymede and Europa that are available leveraging strainlines associated with different patterns. Consequently, it is possible to select a feasible transfer between trajectories exhibiting desired behaviors in the vicinity of each moon. For example, the selected initial conditions for the departure and arrival FTLE maps in Fig.~\ref{fig:availableTransfersFTLEMapsOld} lead to the sample transfer plotted in Fig.~\ref{fig:exampleTransfer} in which the spacecraft, after completing two revolutions of Ganymede and implementing one impulsive maneuver, reaches the interior region of the J--E CR3BP and transits through the vicinity of Europa. Henceforth, the following notation is adopted to illustrate moon-to-moon trajectories in the Jupiter-centered frame (e.g., see Fig.~\ref{fig:exampleTransfer}): instant t\textsubscript{0} denotes the beginning of the transfer; label 0 corresponds to the crossing of the departure section; label 1 indicates the time at which the departure arc intersects the surface of the departure moon SoI; 2 represents the intersection between departure and arrival conics; 3 is the moment in which the arrival conic crosses the surface of the arrival moon SoI; 4 is the time of arrival of the spacecraft at the arrival section; finally, t\textsubscript{f} denotes the end of the transfer. Moreover, the following color scheme is used to link each time label with a specific body: black indicates the spacecraft, orange refers to the departure moon and green corresponds to the arrival moon.


\section{\label{sec:inspectionmaps}Selection of trajectory patterns through inspection maps}
The design of moon--to--moon transfers is greatly aided by combining MMAT access maps with the information provided by FTLE maps, 
as demonstrated in Sect.~\ref{sec:MMATAccessMaps}. Once the Jacobi constant values in the arrival and departure CR3BPs are selected, trajectories that yield a moon--to--moon transfer are identified, assuming that Eq.~\eqref{eq:4.15fromdissertation} is satisfied. Moreover,  transfers with specific mission objectives near each moon are produced. For example, it is possible to determine all the departure conditions that yield a given arrival trajectory (Design 1). Alternatively, one can identify all the feasible arrival conditions for a given  departure trajectory  (Design 2). For illustration purposes, transfers of the two types with Jacobi constant values ${J}_d = 3.00754$ and ${J}_a = 3.00240$, respectively at Ganymede and Europa, and a departure epoch $\theta_{0_{Gan}} = 50^{\circ}$ are here
presented. 
Both design types produce complete transfers dependent on the departure or arrival trajectory, and the other portion of the route (arrival or departure, respectively) is determined. 

\subsection{Design 1: Identifying departure conditions corresponding to a specified arrival trajectory}
\begin{figure}[b!]
\centering
\includegraphics[width=0.75\linewidth]{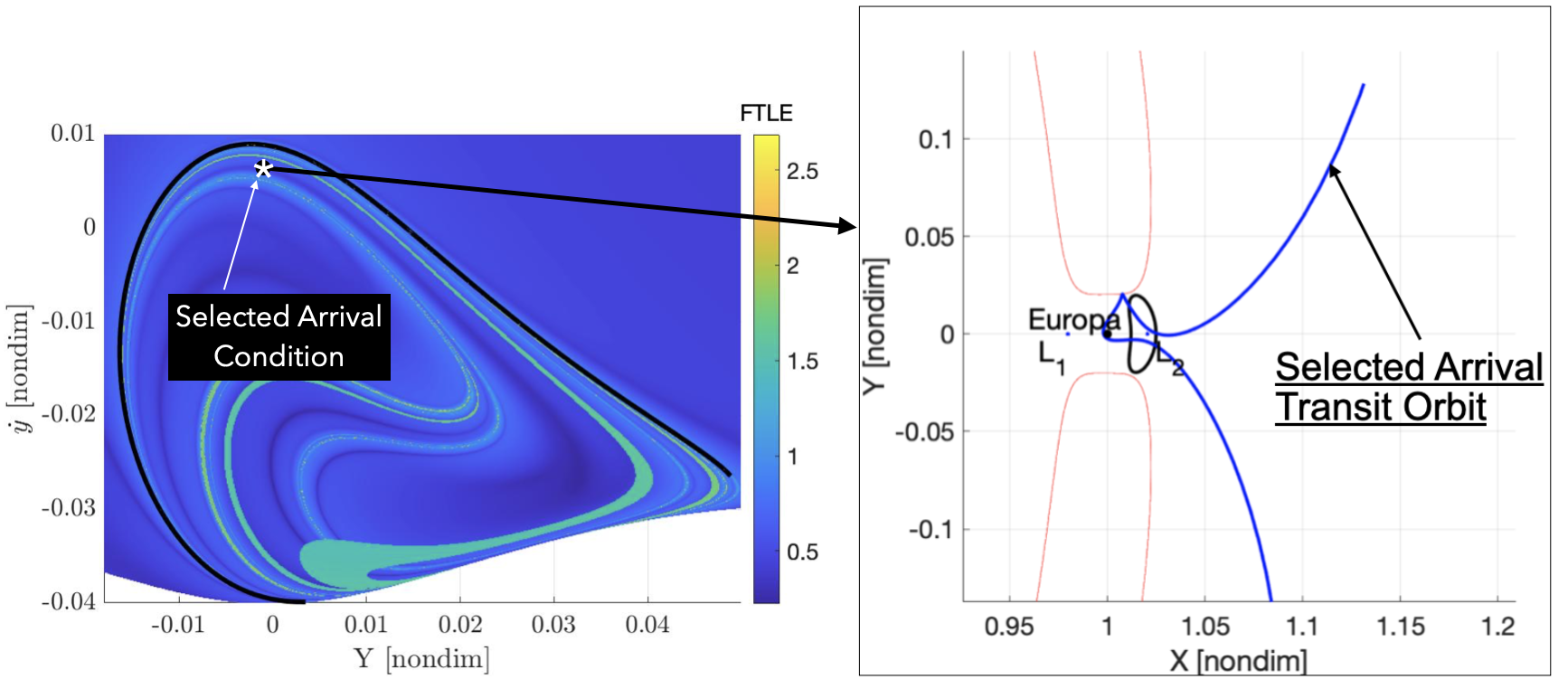}
\caption{Design 1: selected Europa arrival condition.}
\label{fig:europaarrival}
\end{figure}

Assume that the desired arrival trajectory is a close flyby of Europa with approach and departure through the $L_2$ gateway, as depicted in Fig.~\ref{fig:europaarrival}. 
\begin{figure}
\centering
\begin{subfigure}{.4\textwidth}
    
    \includegraphics[width=1\linewidth]{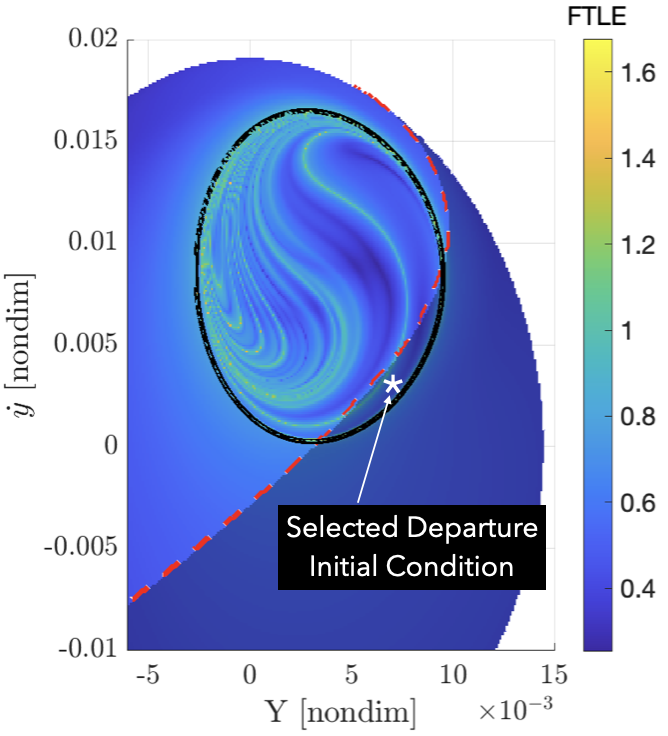}  
    \caption{Ganymede departure FTLE map.}
    \label{SUBFIGURE LABEL 1}
\end{subfigure}
\begin{subfigure}{.4\textwidth}
    \centering
    \includegraphics[width=1\linewidth]{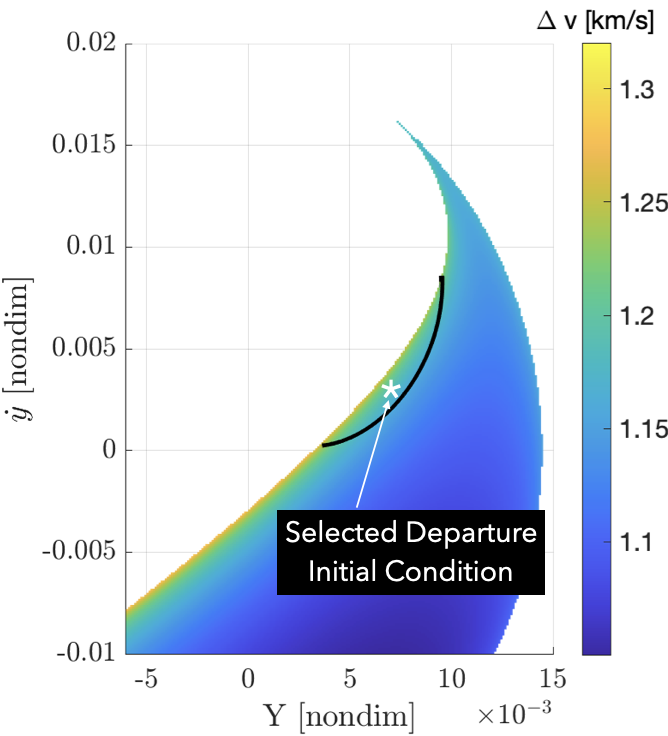}  
    \caption{Total $\boldsymbol{\Delta{v}}$ budget.}
    \label{SUBFIGURE LABEL 2}
\end{subfigure}

\begin{subfigure}{.4\textwidth}
    \centering
    \includegraphics[width=1\linewidth]{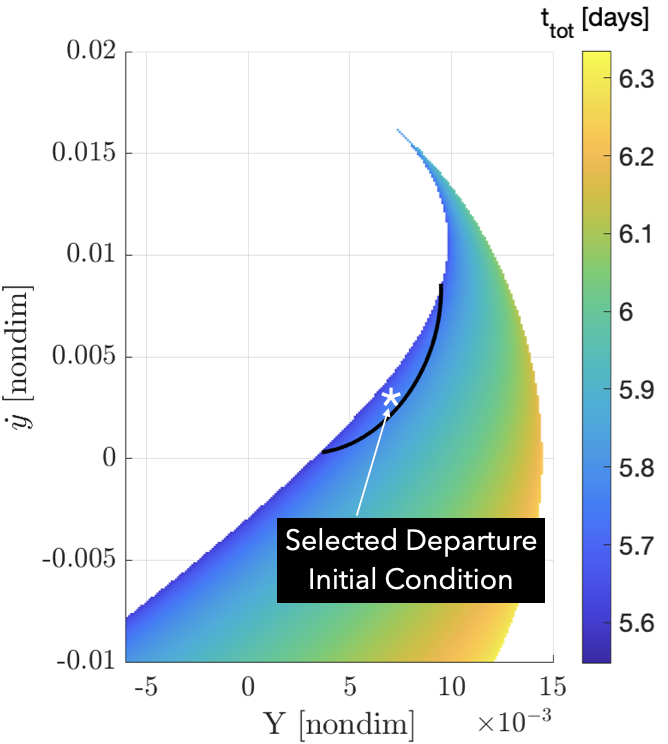}  
    \caption{Total \textbf{time--of--flight}.}
    \label{SUBFIGURE LABEL 3}
\end{subfigure}
\begin{subfigure}{.4\textwidth}
    \centering
    \includegraphics[width=1\linewidth]{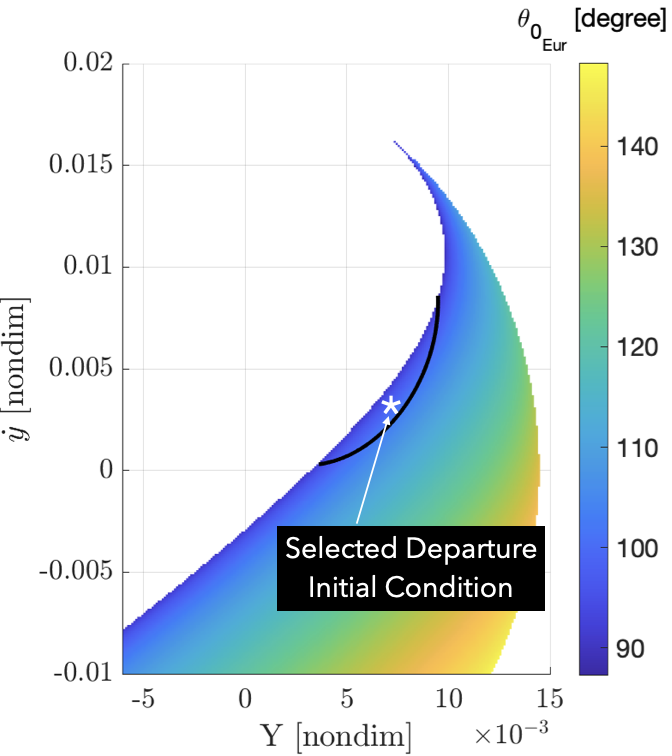}  
    \caption{Phase of Europa at departure.}
    \label{SUBFIGURE LABEL 4}
\end{subfigure}
\caption{Design 1: Ganymede departure FTLE map (a), $\Delta v$ magnitude (b), \textbf{time--of--flight}  (c) and Europa's  phase (d) inspection maps for the Ganymede departure epoch $\boldsymbol{\theta_{0_{Gan}} = 50^{\circ}}$.}
\label{fig:costandphase1}
\end{figure}
The $\Delta{v}$ budget, the total time--of--flight and the relative phase requirements between the moons (Fig.~\ref{fig:costandphase1}) are identified based on the application of the MMAT method to the FTLE map corresponding to departures from Ganymede. Equation  ~\eqref{eq:4.15fromdissertation} is evaluated for all possible transfers with departure conditions from Ganymede matching the selected trajectory arc approaching Europa. Cost and phase inspection maps are employed to obtain cost--effective initial conditions, as apparent in Fig.~\ref{fig:costandphase1}, where the black line corresponds to the unstable manifold and bounds departure options on the maps. A departure trajectory with two revolutions around Ganymede (temporary capture) is selected from among all these opportunities. Eventually, combining the selected departure and arrival trajectories yields the full transfer plotted in Fig.~\ref{fig:finaltransfers}(a). 

\subsection{Design 2: Identifying arrival conditions corresponding to a specified departure trajectory}
\begin{figure}[h]
\centering
\includegraphics[width=0.85\linewidth]{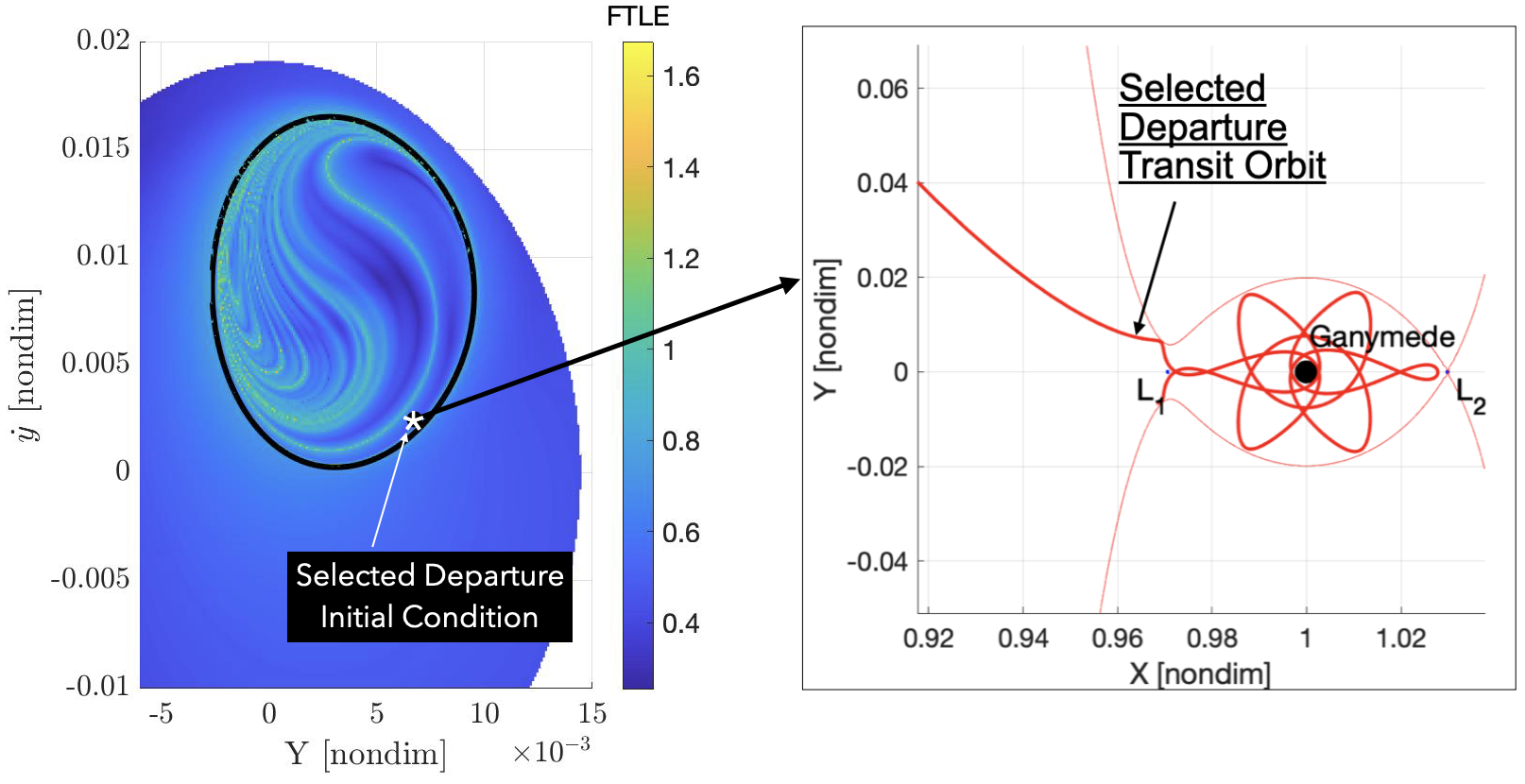}
\caption{Design 2: Selected Ganymede departure condition.}
\label{fig:ganymededeparture}
\end{figure}

\begin{figure}[h]
\centering
\begin{subfigure}{.48\textwidth}
    \includegraphics[width=1\linewidth]{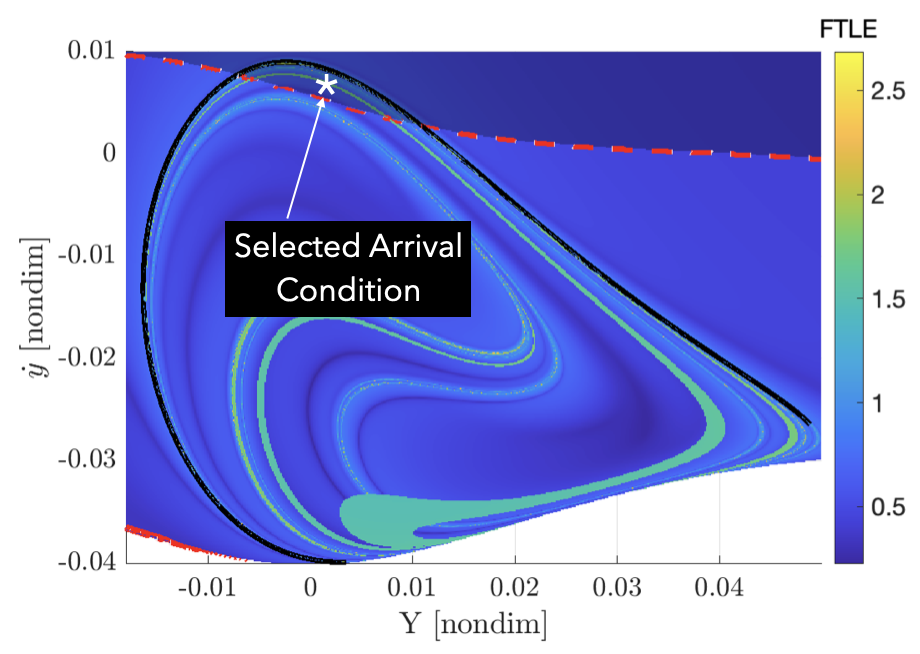}  
    \caption{Europa arrival availability.}
    \label{SUBFIGURE LABEL 1}
\end{subfigure}
\begin{subfigure}{.48\textwidth}
    \centering
    \includegraphics[width=1\linewidth]{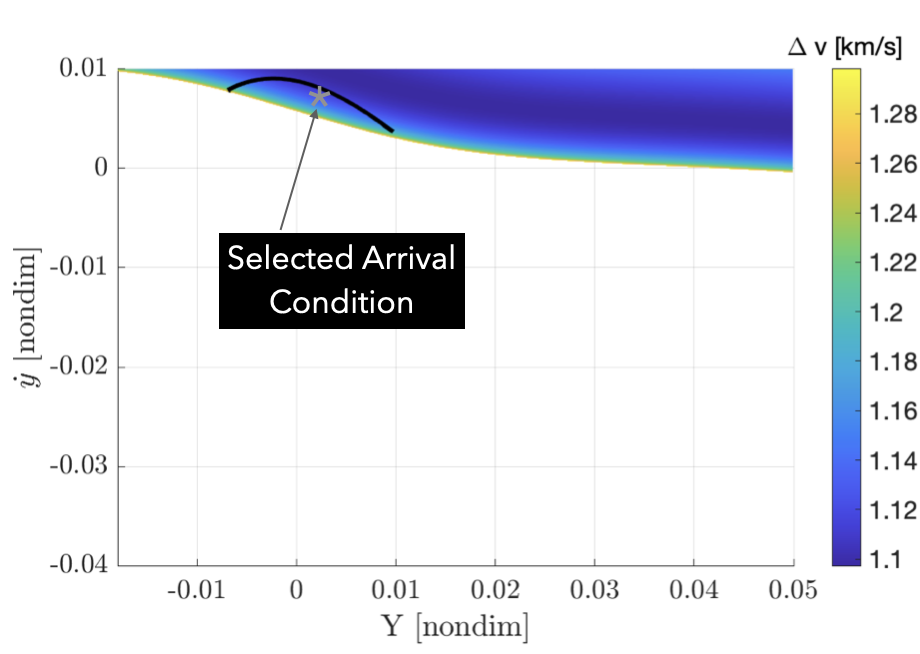}  
    \caption{Total $\boldsymbol{\Delta{v}}$ budget.}
    \label{SUBFIGURE LABEL 2}
\end{subfigure}
\begin{subfigure}{.48\textwidth}
    \centering
    \includegraphics[width=1\linewidth]{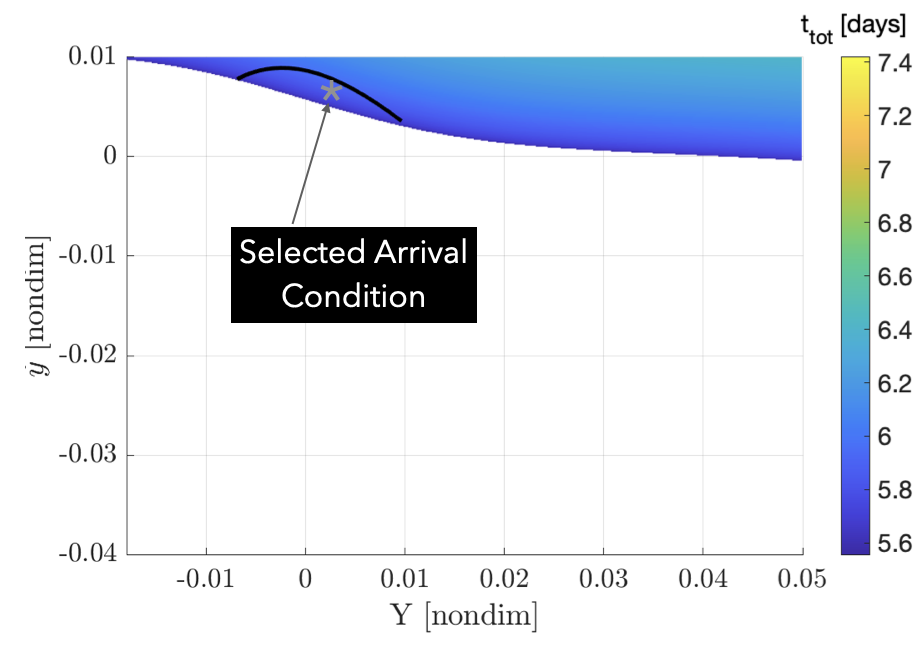}  
    \caption{Total $\boldsymbol{t_{tot}}$ budget.}
    \label{SUBFIGURE LABEL 3}
\end{subfigure}
\begin{subfigure}{.48\textwidth}
    \centering
    \includegraphics[width=1\linewidth]{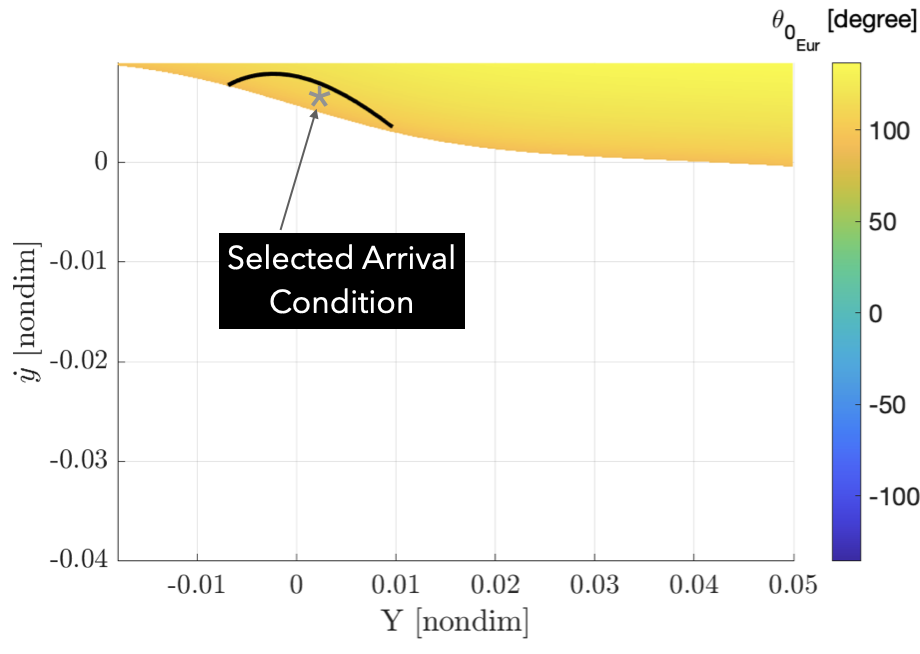}  
    \caption{Europa phase at departure epoch.}
    \label{SUBFIGURE LABEL 4}
\end{subfigure}
\caption{Design 2: Europa arrival FTLE map (a), $\Delta v$ magnitude (b), \textbf{time--of--flight} (c) and Europa's orbital phase (d) inspection maps for the selected Ganymede departure epoch ($\boldsymbol{\theta_{0_{Gan}} = 50^{\circ}}$). }
\label{fig:costandphase2}
\end{figure}

\begin{figure}
\centering
\begin{subfigure}{.85\textwidth}
    \includegraphics[width=1\linewidth]{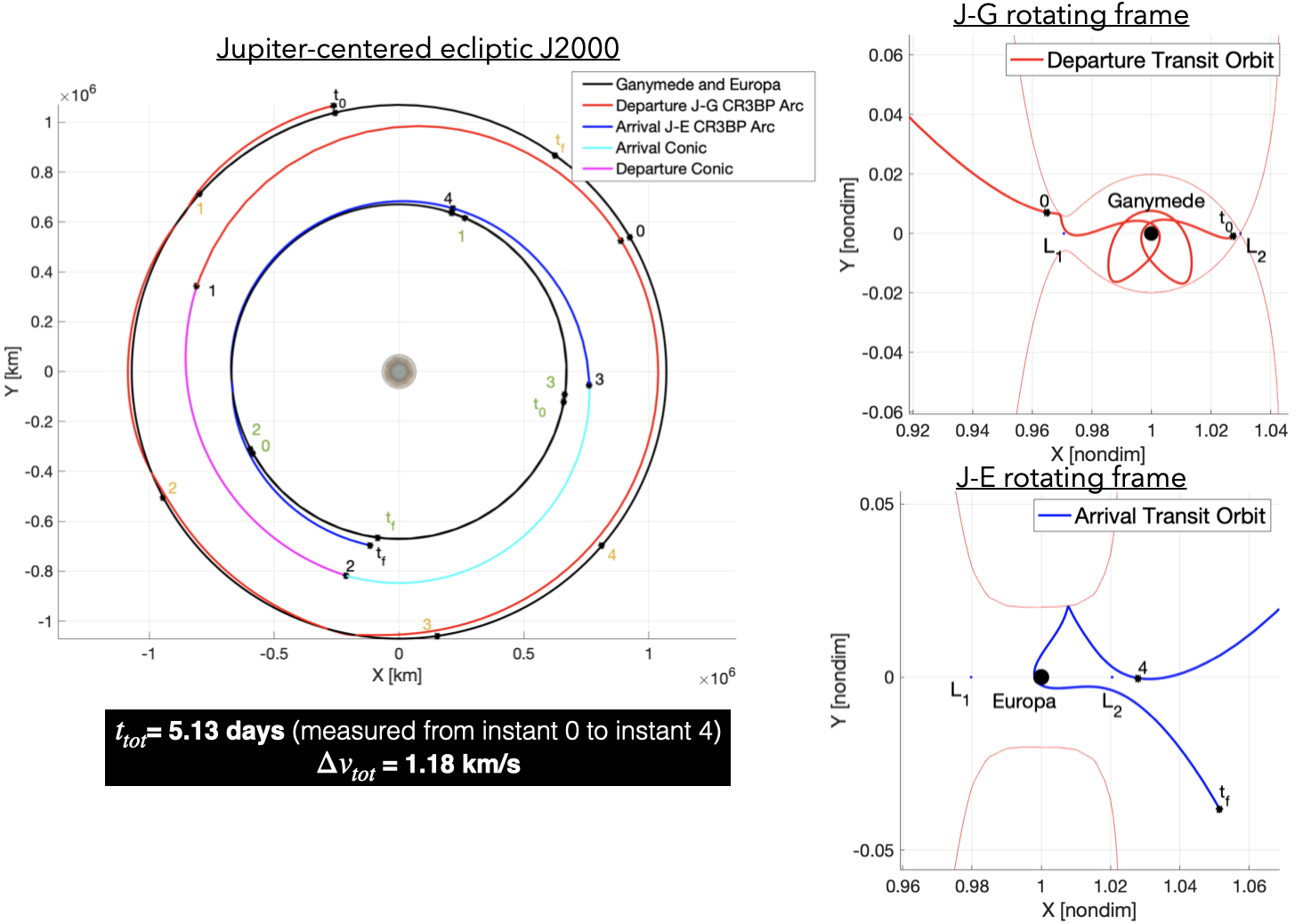}
    \caption{\label{fig:finaltransfer1} Design 1: from temporary capture around Ganymede to a flyby of Europa.}
\end{subfigure}
\begin{subfigure}{.85\textwidth}
    \includegraphics[width=1\linewidth]{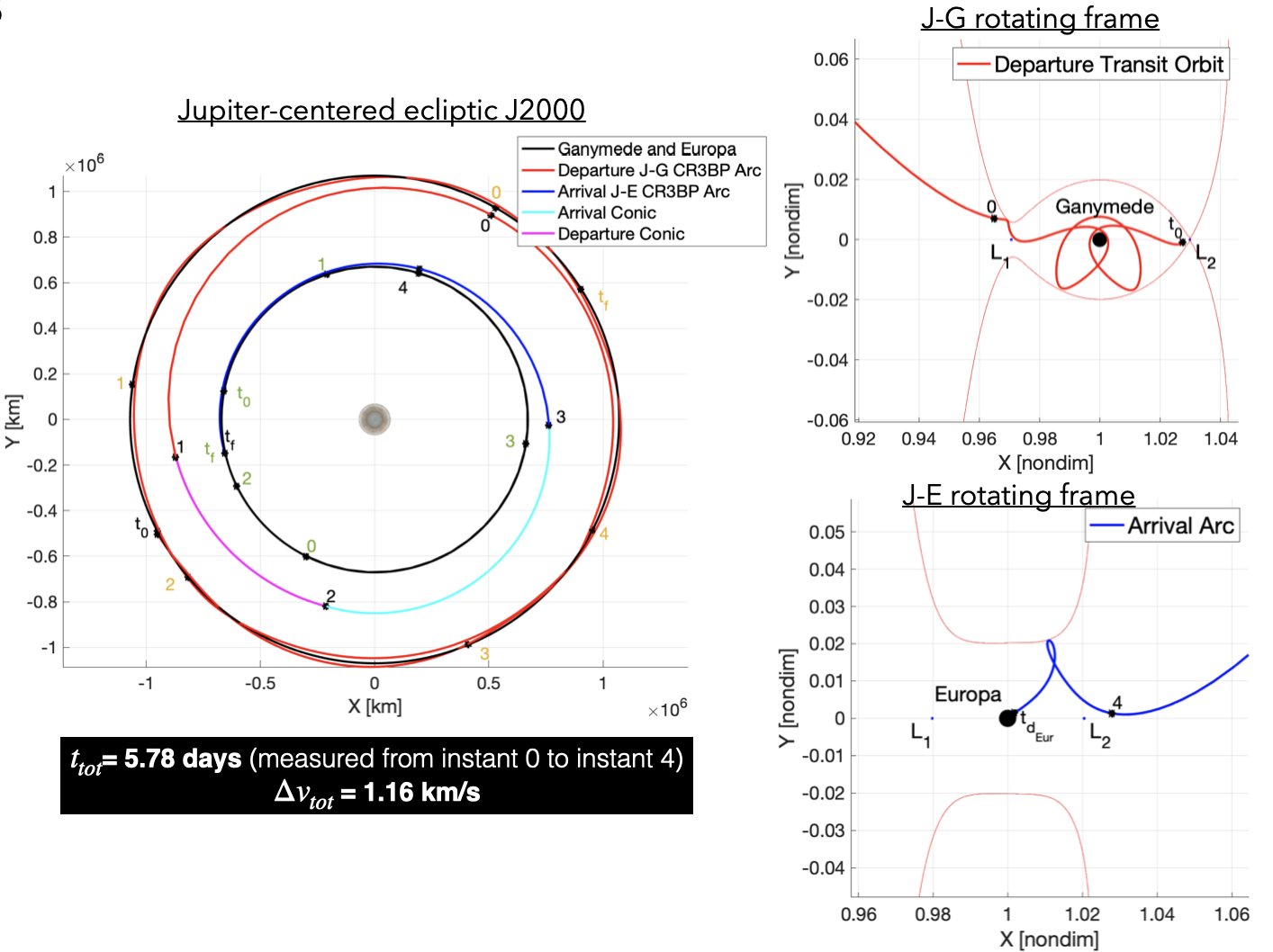}
    \caption{\label{fig:finaltransfer2} Design 2: from temporary capture around Ganymede to landing at Europa.}
\end{subfigure}
\caption{Designs 1 (a) and 2 (b) for planning arrivals to Europa with different endgames.}
\label{fig:finaltransfers}
\end{figure}

In the selected departure trajectory, the spacecraft completes a few revolutions around Ganymede before departure (temporary capture, see Fig.~\ref{fig:ganymededeparture}). By leveraging the MMAT method, all the accessible arrival trajectories near Europa and their associated costs ($\Delta v$) are determined. The initial conditions for arrival are matched to the specified departure arc using Eq.~\eqref{eq:4.15fromdissertation}, and the inspection maps for $\Delta{v}$ budget, total time--of--flight and relative phase requirements between moons (Fig.~\ref{fig:costandphase2}) are produced. For example, the selected solution leads to a landing on the surface of Europa (see Fig.~\ref{fig:finaltransfers}(b)). 

\section{\label{sec:access}Dependence on the Jacobi constant}
\begin{figure}
\hfill{}\centering\includegraphics[width=16cm]{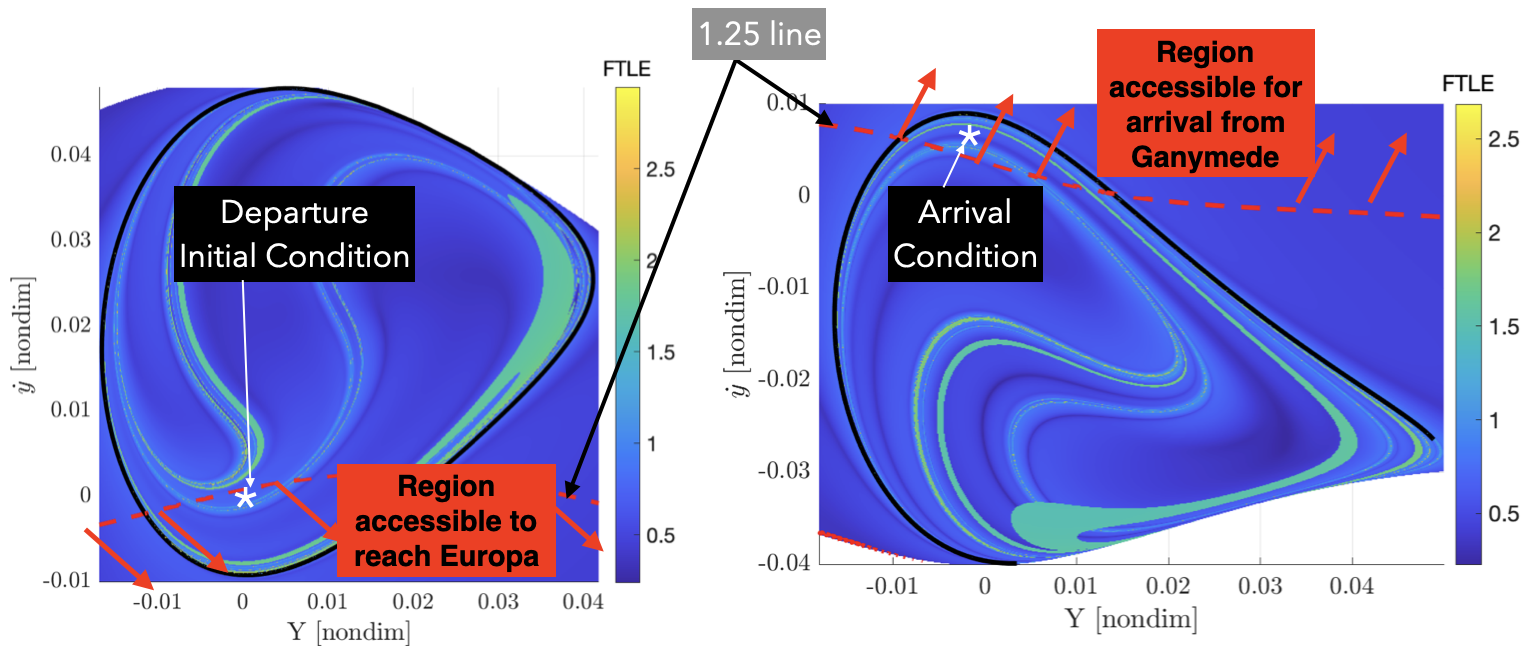}\hfill{}
\caption{\label{fig:availableTransfersFTLEMaps} Available transfers connecting  Ganymede and Europa in the departure and arrival FTLE maps; the 1.25 isoline is used as a reference value for selecting initial conditions.}
\end{figure}

The methodology presented in the previous sections generates different results depending on the Jacobi constant of the departure and arrival trajectories. If the Jacobi constant value for the departure trajectories is decreased ($J_d$ = 3.0061, i.e., the departure energy is higher than in the previous example) while the trajectories approaching Europa remain at $J_a$ = 3.00240, 
the number of initial conditions leading to transfers from Ganymede to Europa increases, as is apparent by comparing  Fig.~\ref{fig:availableTransfersFTLEMapsOld} with Fig.~\ref{fig:availableTransfersFTLEMaps}. Note that an isoline of 1.25 is selected in Fig.~\ref{fig:availableTransfersFTLEMaps} to illustrate the availability of a larger number of initial conditions. Moreover, for the same $J_a$, the lower the departure Jacobi constant ($J_d$) the lower the resulting $\Delta v$ for most available sets of initial conditions and for the same departure epoch. This fact is demonstrated by developing the inspection maps  discussed in Sect. \ref{sec:inspectionmaps} for two departure Jacobi constant values and the same departure epoch. 

\begin{figure}
\hfill{}\centering\includegraphics[width=15.5cm]{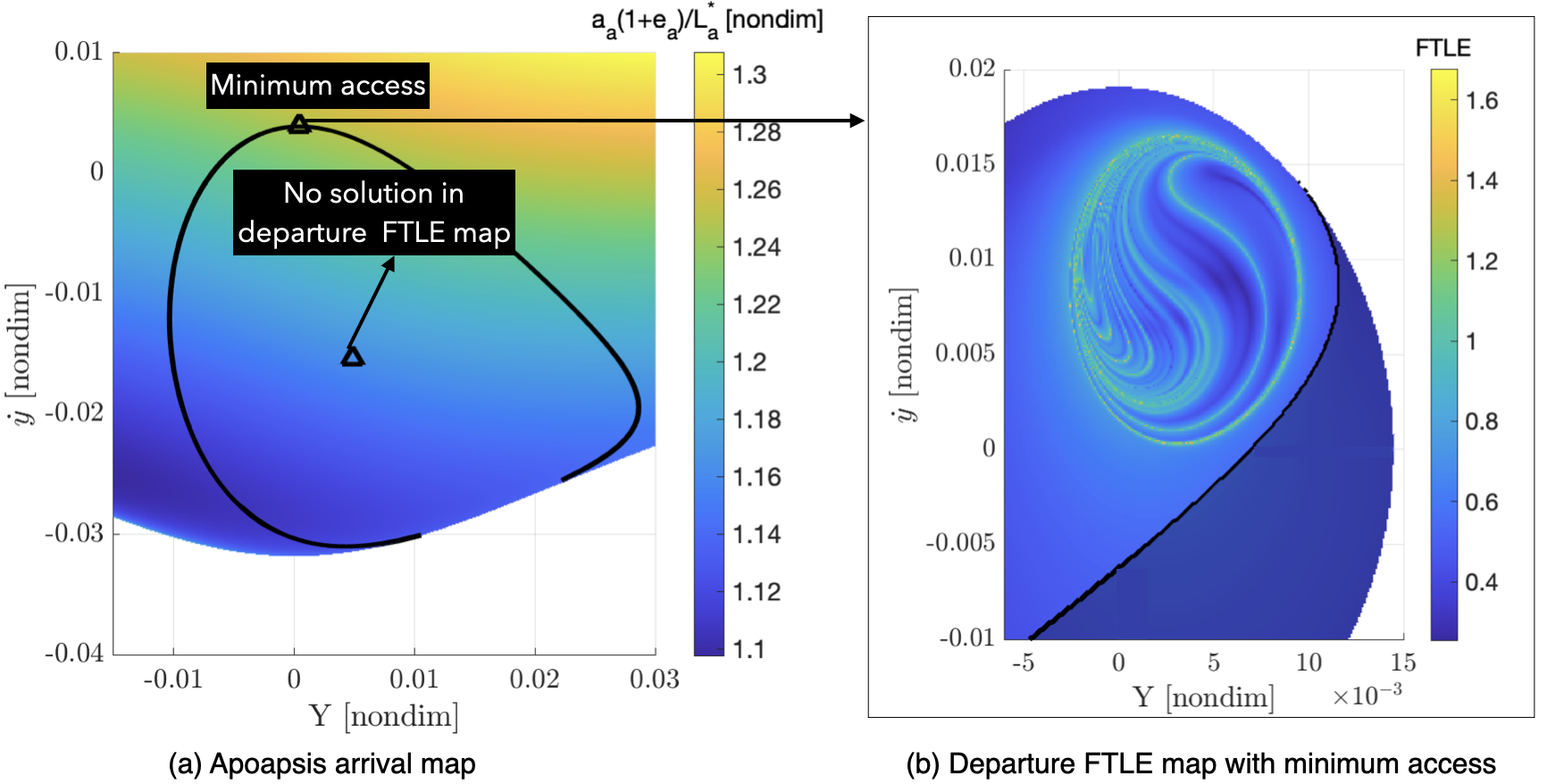}\hfill{}
\caption{\label{fig:reducingJCEuropa} \textbf{Apoapsis arrival map} (a) and departure FTLE map (b) for which Ganymede and Europa cannot be connected for the given combination of energy levels.}
\end{figure}

Now, increase the Jacobi constant level for the arrival trajectories to $J_a$ = 3.0030, i.e., the arrival energy is lower than in the previous two cases, while the trajectories departing Ganymede remain at $J_d$ = 3.00754. If the initial condition 
with the maximum value of $a_{a}(1+e_{a})/L_a^*$ is selected from the apoapsis arrival map, there are no transit orbits departing the Ganymede vicinity that reach the vicinity of Europa (Fig.~\ref{fig:reducingJCEuropa}). 
Hence, the selection of the departure and arrival Jacobi constant values is crucial for the  moon--to--moon design process.

\section{\label{sec:epoch} Dependence on the departure epoch}
Moon--to--moon transfers are epoch--dependent, i.e., the number of connecting trajectories available on the MMAT maps depends on the departure epoch. In the previous examples, the epoch is represented as $\theta_{0_{Gan}}$ (true anomaly of Ganymede at departure). In the following,
two strategies for selecting a departure epoch are discussed: one leverages the arrival hyperbolic stable manifold and the moon--to--moon tides map, whereas the other is based on the departure unstable manifold. For the sake of illustration, the Ganymede--to--Europa scenario is continued.

\subsection{Method A: arrival stable manifold and moon--to--moon tides map}
The apoapsis and periapsis arrival maps respectively represent the apoapsis and periapsis radii of the arrival conics originating
from all the initial conditions that cross the surface of the arrival moon SoI in the arrival FTLE map. Transit orbits near the arrival moon 
are enclosed by the stable hyperbolic invariant manifolds of planar Lyapunov orbits.

\begin{figure}[b!]
\centering
\begin{subfigure}{.45\textwidth}
    \includegraphics[width=1\linewidth]{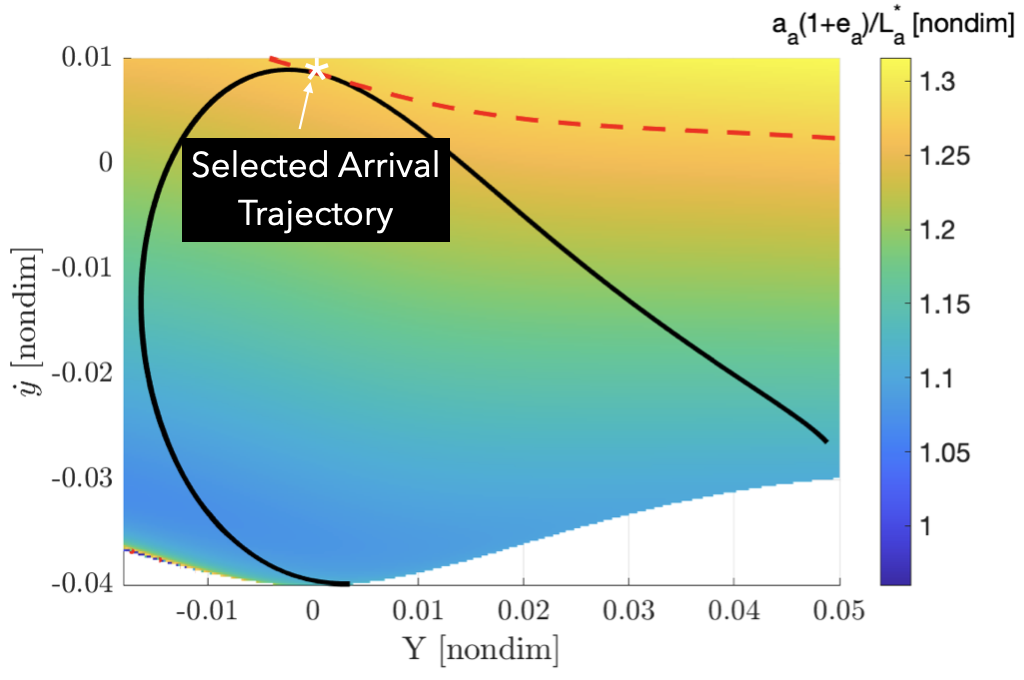}
    \caption{\textbf{Apoapsis arrival map} with 1.2750 isoline overlaid.}
    \label{SUBFIGURE LABEL 2}
\end{subfigure}
\begin{subfigure}{.45\textwidth}
    \includegraphics[width=1\linewidth]{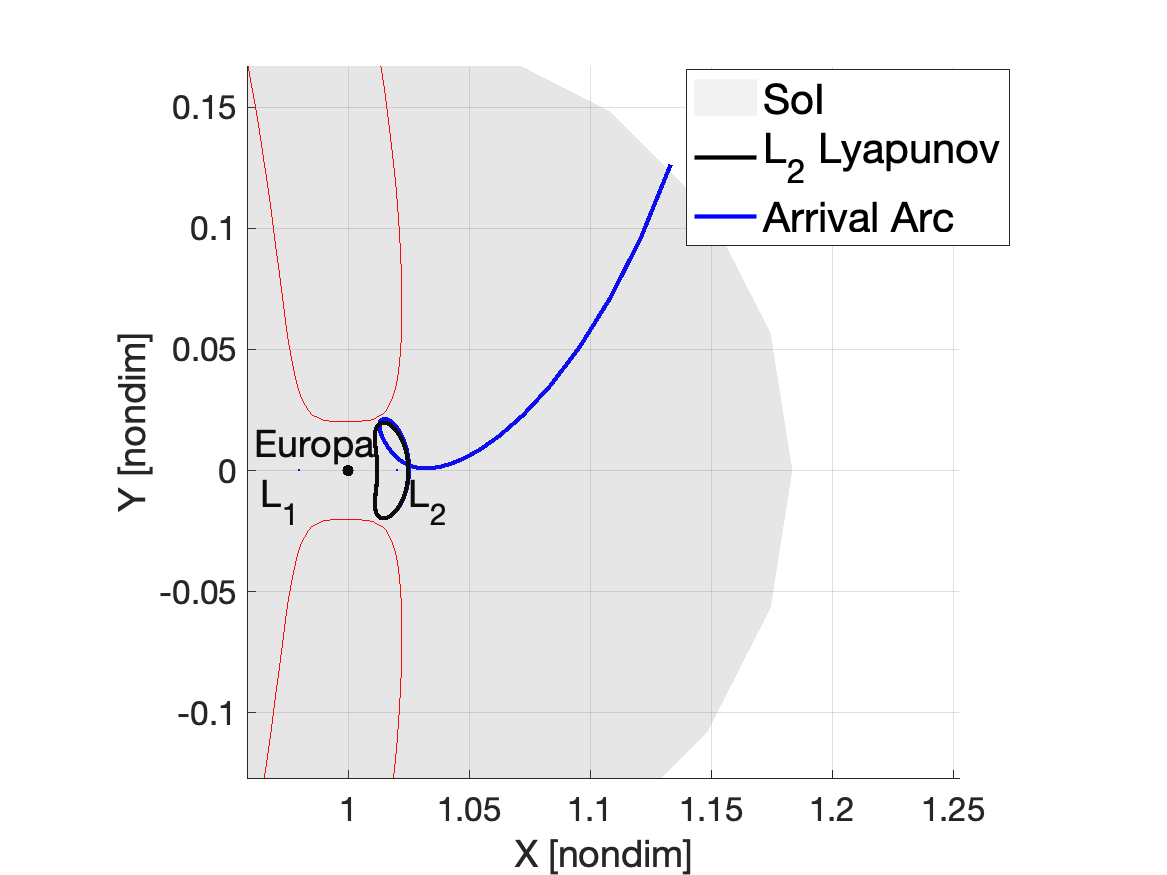}
    \caption{Stable manifold trajectory in the J--E CR3BP resulting in an $\boldsymbol{L_2}$ Lyapunov orbit.}
    \label{SUBFIGURE LABEL 1}
\end{subfigure}
\caption{\textit{Minimum access arrival trajectory} corresponding to the first stable manifold with Ganymede access.}
\label{fig:accesstoGan}
\end{figure}

\begin{figure}[t!]
\centering
\includegraphics[width=0.85\linewidth]{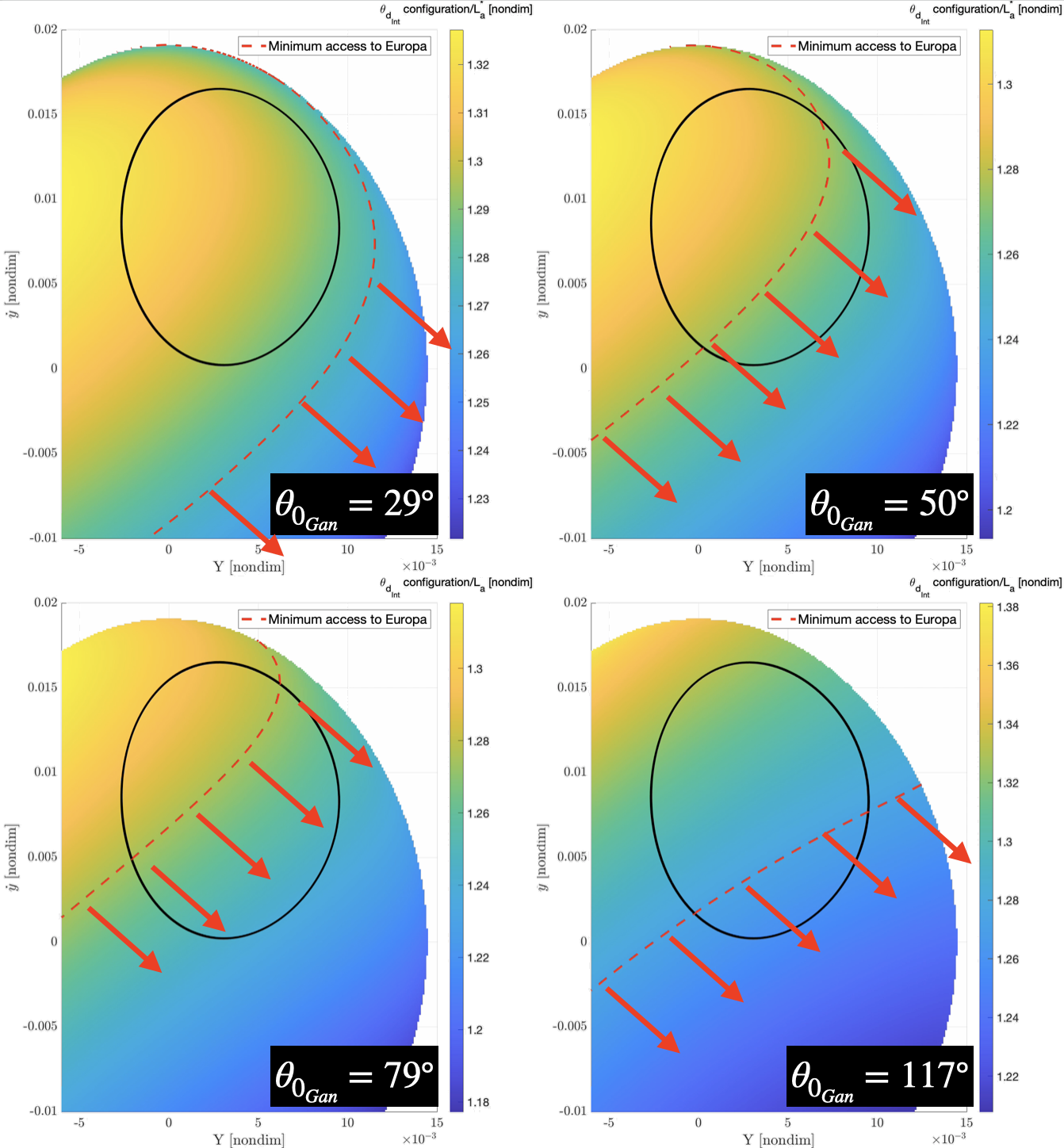}
\caption{\label{fig:FTLEepoch}Europa access variation in moon--to--moon tides map dependent on $\theta_{0_{Gan}}$.}
\end{figure}

To locate the best departure epoch using this method, the stable manifold trajectories of arrival are employed. For an inward journey, the stable manifold trajectory that leads to the maximum $\frac{a_a(1+e_a)}{L_a^*}$ value is chosen. For the opposite journey, a trajectory that provides the minimum value for $\frac{a_a(1-e_a)}{{L_a}^*}$ is selected. From this perspective, the selected trajectory is termed the ``minimum access arrival trajectory'', i.e., the first trajectory that grants access to the arrival moon. As noted, to inspect the available connections from Ganymede to Europa (inward journey), the \textit{minimum access arrival trajectory} is the one that leads to the maximum value of $\frac{a_a(1+e_a)}{L_a^*}$ (first available trajectory to Europa for the selected $J_a$). In this case, the \textit{minimum access arrival trajectory} located at isoline 1.2750 is plotted in Fig.~\ref{fig:accesstoGan} on the corresponding apoapsis arrival map. 
To determine whether transit orbits for a departure from Ganymede and arrival near Europa exist, such an isoline is overlaid onto the moon--to--moon tides map varying $\theta_{0_{Gan}}$. Figure \ref{fig:FTLEepoch} illustrates, for different values of $\theta_{0_{Gan}}$, the trajectories departing the Ganymede vicinity that first deliver access towards Europa.
This procedure demonstrates the variety of possibilities within the range of transfers based on departure epoch. At $\theta_{0_{Gan}} \approx 29^{\circ}$, there is no access to Europa, though departure epochs greater than approximately $35^{\circ}$ do yield transfer options. The number of opportunities increases continuously until a maximum value corresponding to $\theta_{0_{Gan}} \approx 82.5^{\circ}$. Then, the number decreases until $\theta_{0_{Gan}} \approx 130^{\circ}$, where transfers are no longer possible.  

\subsection{Method B: departure unstable manifold}
\begin{figure}
\centering
\includegraphics[width=0.8\linewidth]{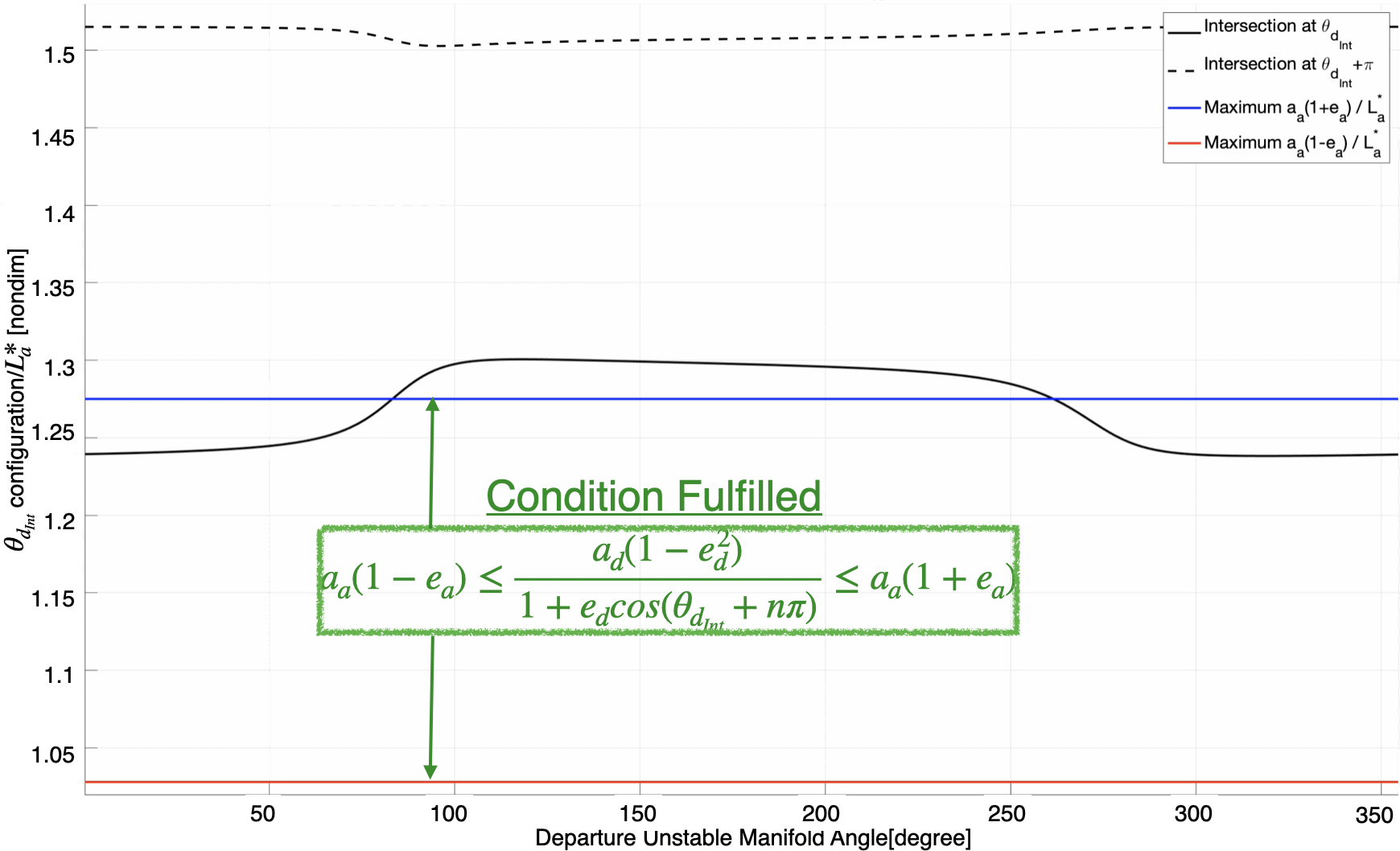}
\caption{\label{fig:variation} Evolution of Eq.  \ref{eq:4.15fromdissertation}  for the entire departure unstable manifold and compared against the \textit{minimum access arrival trajectory}.}
\end{figure}

\begin{figure}
\centering
\includegraphics[width=0.6\linewidth]{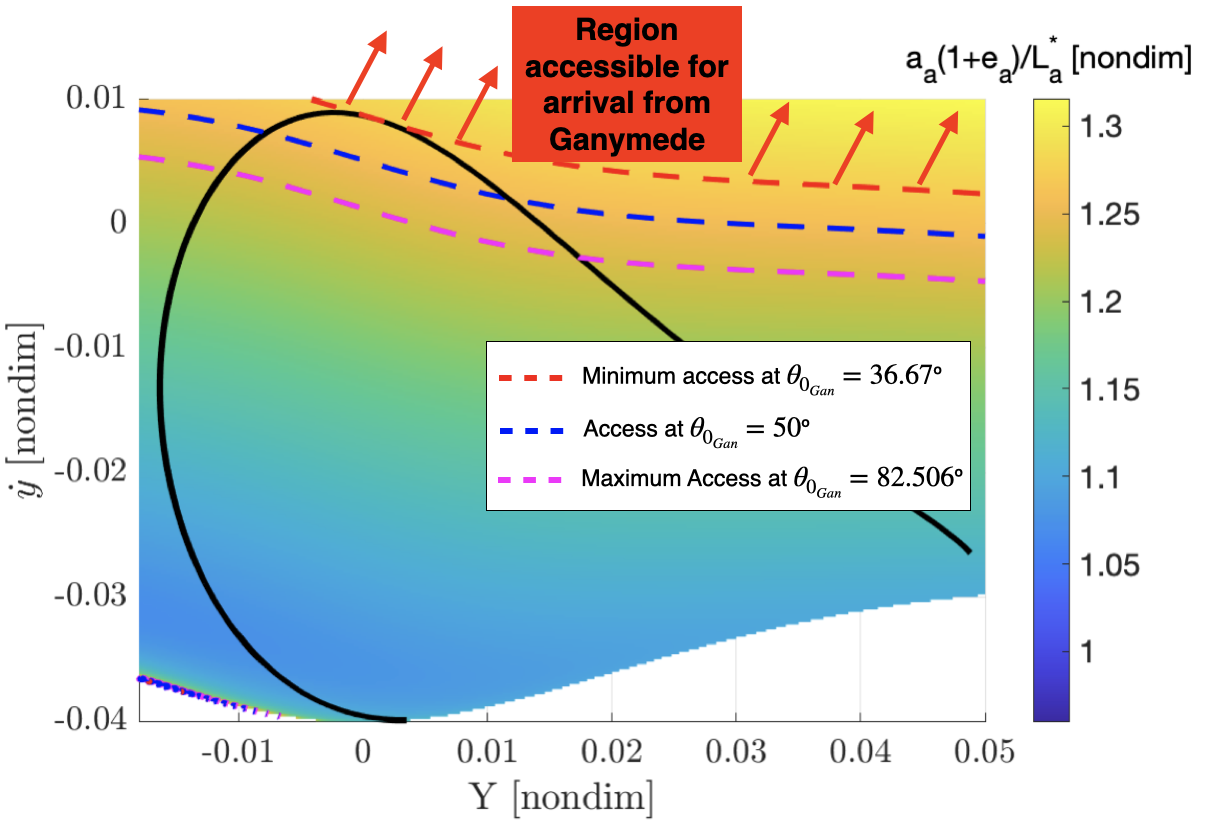}
\caption{\label{fig:access} Apoapsis arrival map depicting accessibility from the first departure trajectory from Ganymede that grants access to Europa.}
\end{figure}

Though more precise, the moon--to--moon tides map approach requires many computations for FTLE maps to discern the various permutations in the departure epochs and identify the date with the largest amount of available transfers. To simplify the process, an alternative strategy utilizing the departure unstable manifold can be employed. The middle term in Eq. \eqref{eq:4.15fromdissertation} is computed for every unstable manifold trajectory propagated towards the SoI of the departure moon over the span of departure epochs from $0^{\circ}$ to $360^{\circ}$. The apoapsis and periapsis for the minimum access arrival trajectory are then compared against the middle term in Eq. \eqref{eq:4.15fromdissertation} for each unstable manifold trajectory. Figure \ref{fig:variation} represent such a comparison based on a departure epoch of ${\theta_{0_{Gan}}=82.5^{\circ}}$: the ``departure unstable manifold angle'' on the horizontal axis corresponds to the location of the departure/arrival arc along the manifold associated with the periodic orbit and measured clockwise from the $\hat{x}$--axis of the J--G rotating frame. The results exhibit significant similarities to those produced from Method A. Utilizing this technique for all $\theta_{0_{Gan}}$ values allows for selection of the unstable manifold trajectories with minimum and maximum values for the middle term in Eq. \eqref{eq:4.15fromdissertation} (inward or outward transfer, respectively). Finally, Fig. \ref{fig:access} illustrates the range of initial conditions upon the arrival FTLE map from the first departure trajectory from Ganymede that grants access to Europa, and the variations depending on $\theta_{0_{Gan}}$. Employing the MMAT technique, a feasibility analysis concerning the budget for $\Delta{v}$ and $t_{tot}$ may, thus, be accomplished for this trajectory. By selecting the lowest $\Delta{v}$, the MMAT access maps only need to be constructed for the given epoch. 

\section{\label{sec:relation} Access dependence between two moons}
\begin{figure}
\centering
\begin{subfigure}{0.9\textwidth}
    \includegraphics[width=1\linewidth]{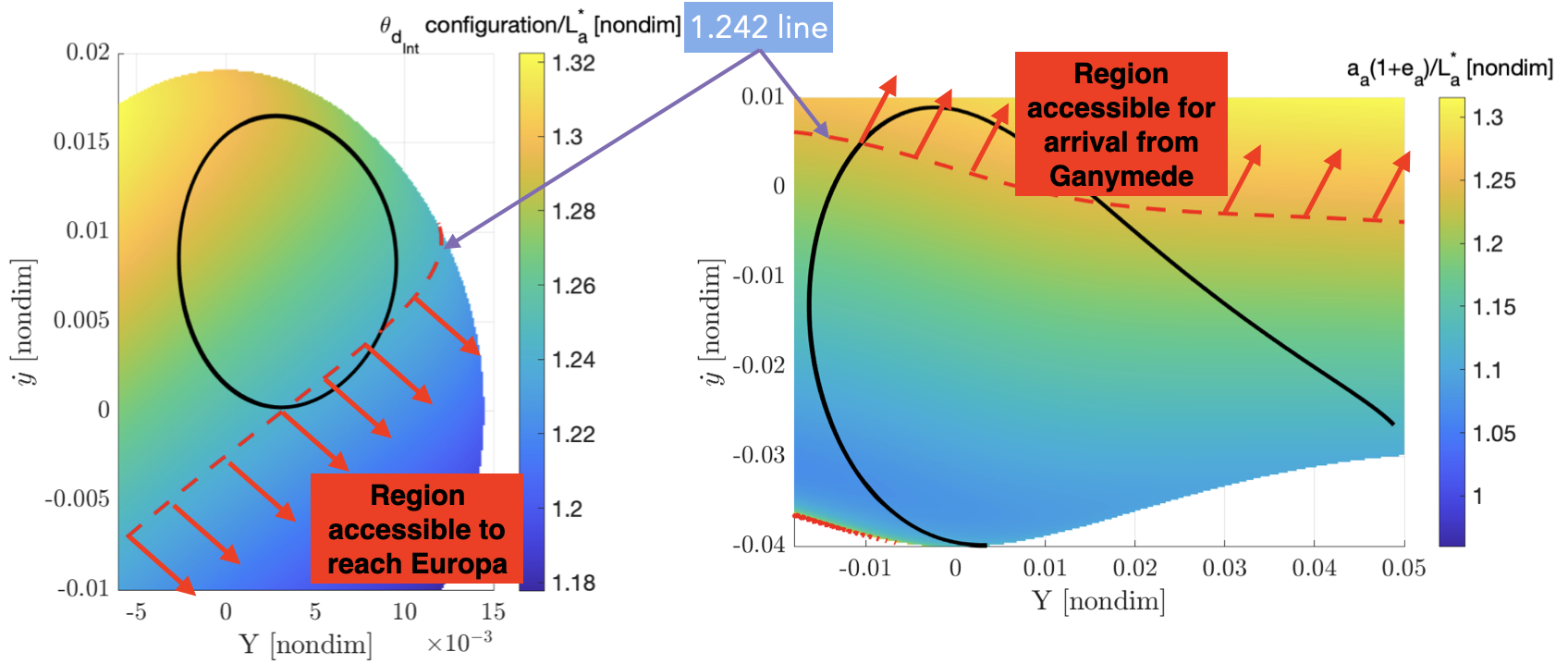}
    \caption{Moon--to--moon tides map overlapped with apoapsis arrival map using 1.242 isoline.}
\end{subfigure}
\begin{subfigure}{0.9\textwidth}
    \includegraphics[width=1\linewidth]{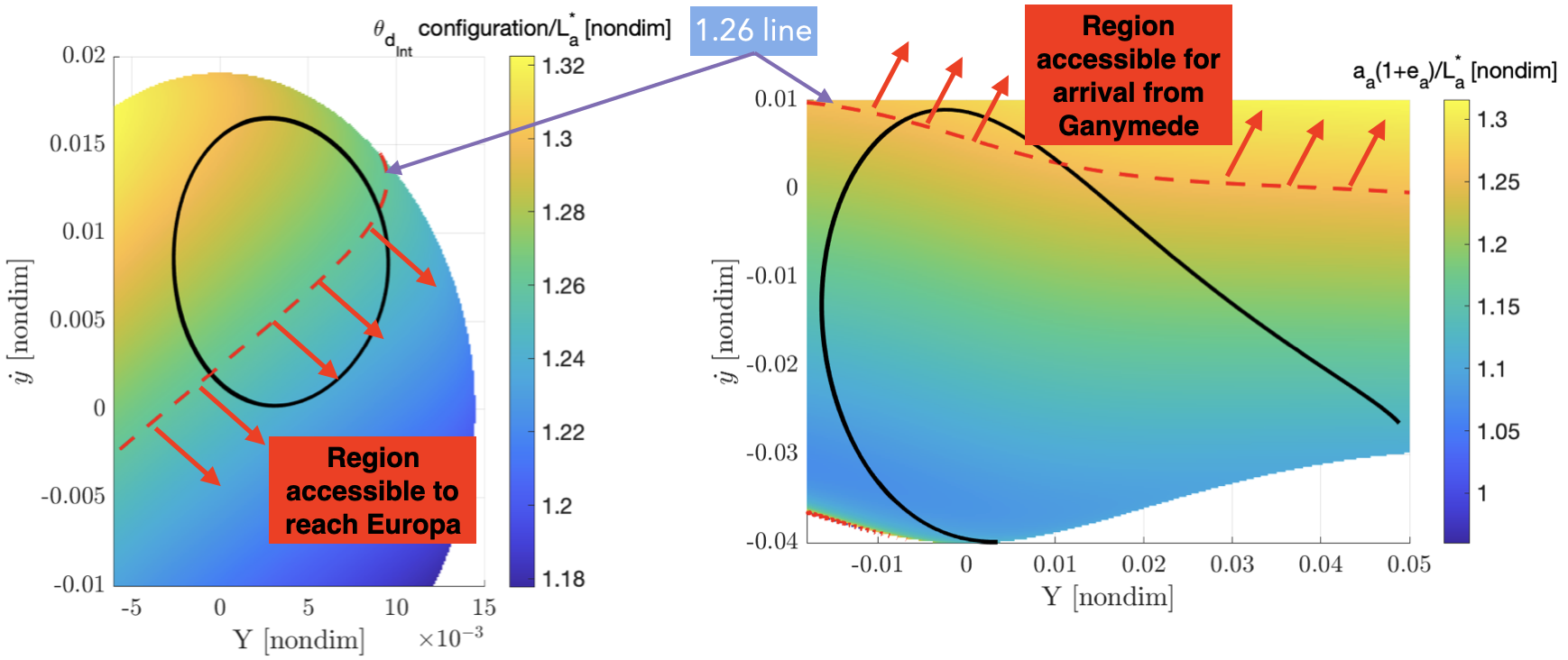}
    \caption{Moon--to--moon tides map overlapped with apoapsis arrival map using 1.26 isoline}
\end{subfigure}
\caption{MMAT maps application to understand access dependence by varying isolines.}
\label{fig:tidesandupper}
\end{figure}

The analysis of the MMAT access maps reveals a relationship concerning the volume of available departure and arrival trajectories that link two moons. For inward transfers, recall that the isolines in the MMAT maps are produced by matching the values of $\frac{{a}_d(1-{{e}_d^2)}}{1+{e}_d\cos({\theta}_{{d}_{Int}})}$ and $a_a(1+e_a)$. For outward transfers, the quantity  $\frac{{a}_d(1-{{e}_d^2)}}{1+{e}_d\cos({\theta}_{{d}_{Int}})}$ is instead set equal to $a_a(1-e_a)$. Consider again the Ganymede--to--Europa case. Extensive experiments with the moon--to--moon tides map and the apoapsis arrival map demonstrate that, for fixed Jacobi constant values at departure and arrival, increasing the isoline value yields a higher number of options in the moon--to--moon tides map, whereas the opposite occurs in the apoapsis arrival map (see Fig.~\ref{fig:tidesandupper}(b)). In contrast, decreasing the isoline value yields more transfer options in the apoapsis arrival maps and fewer in the moon--to--moon tides maps (Fig.~\ref{fig:tidesandupper}(a)):
\begin{enumerate}
\item An increase in the number of transit and unstable manifold trajectories at departure reduces the amount of arrival options;
\item An increase in the number of transit and stable manifold trajectories at arrival reduces the available departure options.
\end{enumerate}
In conclusion, a larger pool of trajectories at departure reduces the set of possible arrival trajectories and vice versa. 
With the moons revolving in their true orbital planes, the above effects are particularly important for mission planning to deliver transfers with single impulsive maneuvers.

\section{\label{sec:additional} Outward versus inward transfers}
\begin{figure}[h!]
\hfill{}\centering\includegraphics[width=15cm]{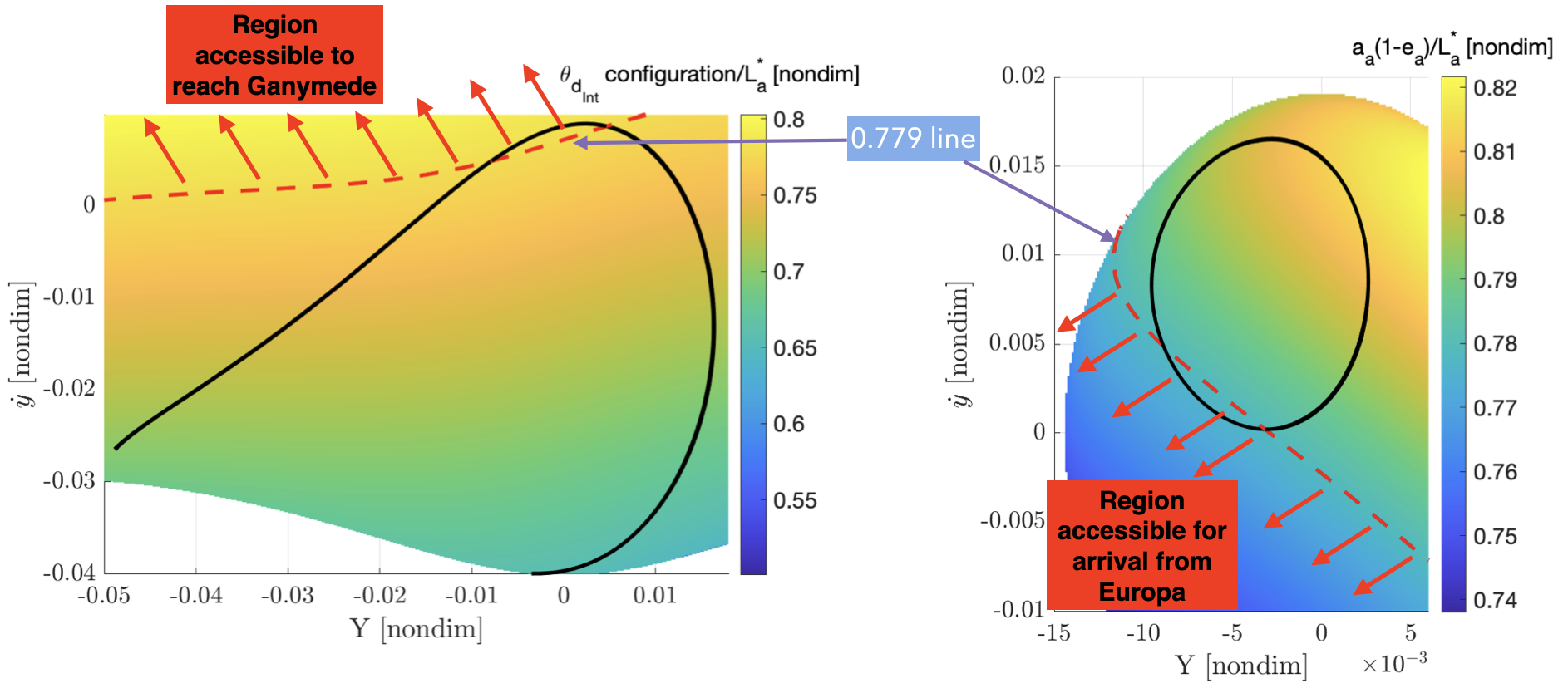}\hfill{}
\caption{\label{fig:europaToGanymede} An overlay between moon--to--moon tides map and periapsis arrival maps for an Europa to Ganymede transfer.}
\end{figure}

The MMAT maps may also be employed in applications to deliver a spacecraft from an inner to an outer moon. As an example, a transfer is designed from Europa to Ganymede with the same Jacobi constant values as in the example of Sect.~\ref{sec:MMATAccessMaps}, i.e., ${J}_d = 3.00240$ (J--E CR3BP) and ${J}_a = 3.00754$ (J--G CR3BP). The transfer options are analyzed using the moon--to--moon tides map in conjunction with the periapsis arrival map. The reference isoline employed  in this case is 0.779:
\begin{align}
\label{eq:TidesConstraintOverlap}
\frac{{a}_d(1-{{e}_d^2)}}{1+{e}_d\cos({\theta}_{{d}_{Int}})}={a}_a(1-{e}_a)=0.779 \; {L}_a^*.
\end{align}
The matching of the two maps is illustrated in Fig.~\ref{fig:europaToGanymede}. The red arrows indicate the initial conditions that 
lead to transfers from Europa to Ganymede. The direction of the arrows is characteristic of the transfer direction (inwards or outwards): for a path from an outer to an inner moon, the arrows are oriented towards the right of the isoline in both FTLE maps (i.e., see Fig.~\ref{fig:MMATMAPS} whereas for transfers from an inner to an outer moon, the arrows are directed to the left of the isoline (see Fig.~\ref{fig:europaToGanymede}).

\section{\label{sec:conclusion} Conclusions}
This contribution introduces the use of finite--time Lyapunov exponent (FTLE) maps for the selection of desired motion patterns in the vicinity of the departure and destination bodies within moon--to--moon transfers. 
The options that can be identified include gravitational captures, lunar surface impacts (or landings), transits through the vicinity of the moons, and insertion into libration point periodic orbits. The FTLE  maps have been blended with the Moon--to--Moon Analytical Transfer (MMAT) method and applied to the design of direct 3D single--impulse transfers between  two Jovian moons, i.e., Ganymede and Europa. The technique offers a very simple 
visualization of different types of motion through so-called access maps. 
Their computation, constructed using a combination of MATLAB and Java scripts, is relatively fast, taking approximately 10 minutes on a Macbook Pro (2.3 GHz Intel Core i9, 16 GB RAM) per map at a typical resolution of $10^{-4}$ in $y$ and $\dot{y}$ in normalized CR3BP units.  This performance can be improved by 25-50\% if only the calculation of the inner portion of the map (i.e., internal to the manifold) is carried out.

The number of solutions, their type and performance (i.e., $\Delta v$ and time of flight) depend on the launch date and the Jacobi constant of the departure and destination orbits, hence the technique is suitable for trade off and optimization studies. Moreover, if the spacecraft performs multiple revolutions around the planet on the selected Keplerian orbits, more phasing options become available between departure and destination bodies for the same $\Delta v$.

Although the case analysed throughout the paper is a simple impulsive connection between planet--centered ellipses originating from libration point periodic orbits at the departure and arrival moons, it constitutes a solid foundation for the development of the methodology. 
Firstly, the direct transfer can provide the initial guess for the design of low-thrust trajectories, in this way alleviating the
mass budget corresponding to the large associated $\Delta v$ values. Secondly, blending FTLE maps with the MMAT technique 
has served the purpose of illustrating the method and maintaining the focus of the discussion on its properties and benefits, rather than on the inter--moon transfer design methodology. 
As a matter of fact, FTLE maps can assist in the design of the begin game and end game of a moon--to--moon tour constructed on intermediate gravity assists and resonant orbits: the information provided by FTLE maps allows an assessment of the specific type of trajectory connecting an initial condition in a low moon orbit with the first resonance, a valuable possibility when shaping the science phase of a mission. It is interesting to observe that
among the different motion types illustrated in this paper, takeoffs and landings are begin games and end games themselves if the moon is in tidal lock with the planet (it is the case of the Galilean moons as well as the Inner Large Moons of Saturn) because, in this case, the positions of the launch and landing sites in the planet--moon synodic barycentric reference frame can be straightforwardly mapped to geographical coordinates. 
In conclusion, the proposed FTLE maps are useful to characterize the space around the moon and study the available options within the envelope of a loose capture, which is the intermediate stage of any low--energy transfer involving approach and escape.

\section*{Acknowledgments}
Assistance from colleagues in the Multi--Body Dynamics Research group at Purdue University is acknowledged as is the support from the Purdue University School of Aeronautics and Astronautics and College of Engineering including access to the Rune and Barbara Eliasen Visualization Laboratory. D. Canales is currently supported by the Department of Aerospace Engineering in Embry--Riddle Aeronautical University. E. Fantino acknowledges Khalifa University's internal grant 8474000413/CIRA--2021--065 and projects PID2020--112576GB--C21 and PID2021--123968NB--100 awarded by the Spanish Ministry of Science and Innovation.  

\bibliography{references}

\end{document}